# HOTSPOT-YOLO: A Lightweight Deep Learning Attention-Driven Model for Detecting Thermal Anomalies in Drone-Based Solar Photovoltaic Inspections

Mahmoud Dhimish[1,*]

[1] Technical University of Denmark, Department of Photonics Engineering, Roskilde, Sjælland, 4000, Denmark

[*]Corresponding Author: Mahmoud Dhimish (E-mail: mahdh@dtu.dk)

## Abstract

Thermal anomaly detection in solar photovoltaic (PV) systems is essential for ensuring operational efficiency and reducing maintenance costs. In this study, we developed and named HOTSPOT-YOLO, a lightweight artificial intelligence (AI) model that integrates an efficient convolutional neural network backbone and attention mechanisms to improve object detection. This model is specifically designed for drone-based thermal inspections of PV systems, addressing the unique challenges of detecting small and subtle thermal anomalies, such as hotspots and defective modules, while maintaining real-time performance. Experimental results demonstrate a mean average precision of 90.8%, reflecting a significant improvement over baseline object detection models. With a reduced computational load and robustness under diverse environmental conditions, HOTSPOT-YOLO offers a scalable and reliable solution for large-scale PV inspections. This work highlights the integration of advanced AI techniques with practical engineering applications, revolutionizing automated fault detection in renewable energy systems.

**Keywords:** Thermal anomaly detection; Solar photovoltaic inspections; Object detection; Convolutional neural network.

# 1. Introduction

The advancement of renewable energy technologies, particularly solar photovoltaics (PV), has gained significant attention due to the global push toward sustainability and carbon neutrality. Ensuring the operational efficiency and reliability of solar PV systems is crucial for maximizing energy output and reducing maintenance costs. Thermal infrared (IR) imaging has emerged as a pivotal technology for inspecting and diagnosing PV systems [1-3], offering a non-invasive and real-time method to identify anomalies such as hotspots [4,5], defective modules [6], and mismatched cells [7]. These thermal anomalies often signify critical issues, including degradation, shading effects, or electrical faults, which can compromise the performance and safety of PV installations.

Thermal IR imaging is extensively used in the solar industry for monitoring PV systems, particularly in large-scale installations [8]. Its ability to visualize temperature variations across PV arrays enables early detection of faults that are otherwise invisible under standard visual inspections. These inspections are typically conducted using drones equipped with thermal cameras, allowing rapid coverage of expansive areas. However, the analysis of thermal images presents unique challenges [9,10] due to variations in environmental conditions, resolution constraints, and the subtle nature of some thermal anomalies.

Traditional methods for thermal image analysis often rely on manual inspection or rule-based algorithms, which are time-consuming and prone to errors, particularly in large-scale deployments. Moreover, these approaches struggle to adapt to the diverse and complex thermal patterns encountered in real-world scenarios. This limitation underscores the need for data-driven approaches capable of automating the detection process while maintaining high accuracy and robustness.

Recent advancements in artificial intelligence (AI) and deep learning have revolutionized object detection tasks, providing powerful tools for automating the analysis of complex datasets. Convolutional Neural Networks (CNNs) [11-13] have demonstrated exceptional performance in extracting hierarchical features from images, enabling precise detection and localization of objects. Specifically, object detection models such as the You Only Look Once (YOLO) [14,15] family have gained prominence for their ability to balance accuracy and computational efficiency, making them ideal for real-time applications like drone-based PV inspections. Despite their success, applying general-purpose models like YOLO to specialized domains such as thermal imaging requires architectural modifications and domain-specific optimizations [16]. The thermal domain presents unique challenges, including low contrast, noise, and the presence of subtle anomalies that demand enhanced feature extraction capabilities.

The YOLO framework is a cornerstone of modern object detection, renowned for its speed and accuracy. Unlike traditional multi-stage detection pipelines, YOLO treats object detection as a single regression problem [17], predicting bounding boxes and class probabilities directly from the input image. YOLOv11 [18], the latest iteration, incorporates advanced techniques such as multi-scale feature aggregation and anchor-free mechanisms to further improve detection performance.

To tailor YOLO for thermal anomaly detection in PV systems, several enhancements are necessary. Incorporating lightweight backbones, such as EfficientNet [19], optimizes feature extraction for thermal images, balancing accuracy and computational demands. Additionally, attention mechanisms like Squeeze-and-Excitation (SE) [20,21] blocks enable the model to focus on thermally significant regions, improving the detection of small and subtle anomalies. These architectural modifications ensure that the adapted model can address the specific requirements of drone-based thermal imaging, such as real-time processing and high precision.

To further enhance the detection capabilities in thermal imaging, multi-scale feature aggregation [22] plays a critical role. Thermal anomalies, especially in solar PV systems, can vary significantly in size and intensity, making it essential for the model to capture both fine-grained local features and broader contextual patterns. Multi-scale feature aggregation achieves this by integrating low, medium, and high-level features, ensuring a comprehensive representation of the input image [23]. This hierarchical approach allows the detection model to simultaneously identify small hotspots and larger defective regions, thereby improving the robustness and accuracy of anomaly detection under varying conditions.

Another important consideration is the post-processing of detection outputs. Thermal imagery often contains overlapping or redundant detections, particularly in noisy or cluttered environments. Techniques like Non-Maximum Suppression (NMS) [24,25] are essential for refining the final predictions by eliminating duplicate bounding boxes and retaining only the most confident detections. This step is crucial for maintaining precision and clarity in the results, especially when dealing with large-scale PV arrays where overlapping anomalies may occur.

In this study, we introduce HOTSPOT-YOLO, a lightweight and attention-driven adaptation of YOLOv11 specifically tailored for detecting thermal anomalies in PV systems. By incorporating an EfficientNet backbone, the model optimizes feature extraction, achieving a remarkable balance between detection accuracy and computational efficiency. This adaptation is critical for handling drone-based inspections, where real-time processing is a necessity. Additionally, the integration of SE attention mechanisms enables HOTSPOT-YOLO to focus on thermally significant regions, significantly enhancing its ability to detect subtle anomalies, such as small hotspots, with precision. The proposed model achieves a mean average precision of 90.8%, which is a notable improvement of 5.9% compared to the baseline YOLOv11 model. Furthermore, HOTSPOT-YOLO reduces the number of parameters by 2.12 million, reducing the computational overhead while maintaining robust performance across diverse imaging conditions, including variations in brightness, contrast, and thermal noise. This reduction makes it not only highly effective but also deployable on resource-constrained platforms, such as drones. The architecture's ability to adapt to high-noise environments and isolate meaningful anomalies demonstrates its potential for scalable applications in large-scale PV inspections, making it a practical and efficient solution for addressing the growing demand for automated solar monitoring systems.

## 2. Methodology

## 2.1 Description of HOTSPOT-YOLO Model

YOLOv11 is the latest iteration of the "You Only Look Once" family of object detection models [18], renowned for its state-of-the-art performance in real-time object detection tasks. Building on the legacy of its predecessors, YOLOv11 represents a significant advancement in detection accuracy, computational efficiency, and scalability, making it one of the most advanced models available for a wide range of applications. However, while YOLOv11 excels in generic object detection tasks, adapting it to specialized domains, such as thermal anomaly detection in solar PV systems, requires careful architectural enhancements and domain-specific optimizations.

To address this need, we propose the HOTSPOT-YOLO model (presented in Figure 1), an enhanced version of YOLOv11 tailored for drone-based thermography. HOTSPOT-YOLO incorporates a lightweight EfficientNet backbone to optimize feature extraction, offering a superior balance between accuracy and computational efficiency, which is critical for processing thermal imagery in real-time. Furthermore, we integrated SE attention mechanisms into the architecture to enable the model to focus on the most thermally significant regions in the images, such as hotspots or defects, while suppressing irrelevant background information. These modifications were introduced to improve the model's detection accuracy for small and subtle thermal anomalies that are often difficult to detect with conventional approaches. In addition, by maintaining the core real-time capabilities of YOLOv11 and optimizing its computational demands, HOTSPOT-YOLO is designed to be deployable on drones for large-scale solar PV inspections. These enhancements ensure that the model not only achieves high detection performance but also remains practical and efficient for real-world applications.

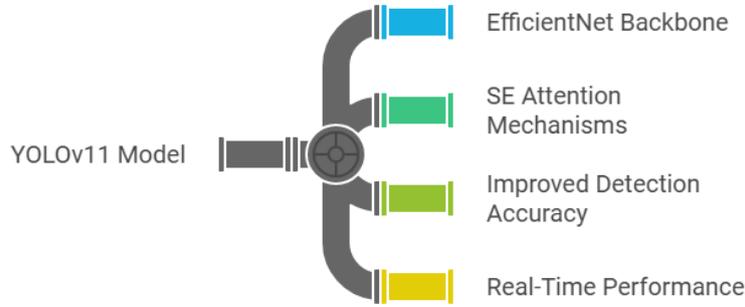

**Figure 1.** Conceptual overview of the HOTSPOT-YOLO model. The model builds upon YOLOv11 with significant enhancements, including a lightweight EfficientNet backbone for feature extraction, Squeeze-and-Excitation attention mechanisms to focus on critical regions, improved detection accuracy, and real-time performance optimized for drone-based thermal anomaly detection.

## 2.2 Detailed Operation of HOTSPOT-YOLO Model

The proposed HOTSPOT-YOLO architecture, illustrated in Figure 2, operates by systematically transforming input thermal imagery into bounding box predictions and class labels that identify thermal anomalies in solar PV systems. The process is structured into multiple computational stages, which are seamlessly integrated to maximize efficiency and accuracy. Each stage is described below, along with its corresponding mathematical formulations.

The input images, captured through drone thermography, are resized to a fixed resolution of $H \times W = 640 \times 640$ and normalized to ensure consistent intensity distribution. Let $X \in R^{H \times W \times 3}$ represent the pre-processed input image. The EfficientNet backbone is responsible for extracting hierarchical features from the input image [26]. The backbone comprises a series of depthwise separable convolutions, denoted as (1).

$$F_{i+1} = \sigma(Conv_{depthwide}(F_i) * Conv_{pointwise}(F_i)) \tag{1}$$

where $F_i$ is the input feature map at stage $i$, $\sigma$ represents the activation function (ReLU), and $*$ denotes the convolution operation. To enhance the feature extraction process, Squeeze-and-Excitation blocks are integrated into the backbone. The SE mechanism recalibrates feature maps by first applying Global Average Pooling (GAP) [27] as in (2). One of the key advantages of using GAP is its ability to retain the most critical information while discarding spatial details, enabling the model to learn channel-wise dependencies rather than spatial relationships. This simplification ensures that the SE mechanism can focus on recalibrating the importance of feature maps based on their contribution to the task, such as detecting small and subtle thermal anomalies in solar PV systems. Additionally, GAP eliminates the need for fully connected layers at the classification stage [28], reducing the model's complexity and risk of overfitting.

$$s_c = \frac{1}{H \times W} \sum_{h=1}^{H} \sum_{w=1}^{W} F_c(h, w) \tag{2}$$

where $s_c$ is the scalar summary of channel $c$. This is followed by fully connected layers that learn channel-wise importance, and the recalibrated features, $F_c^{SE}$, are multiplied back into the original feature maps as expressed by (3).

$$F_c^{SE} = F_c \cdot \sigma(W_2 \cdot \text{ReLU}(W_1 \cdot s_c)) \qquad (3)$$

where $W_1$ and $W_2$ are the learnable weights of the fully connected layers.

The recalibration process detailed in Equation (3) reflects the core functionality of the Squeeze-and-Excitation block, where channel-wise attention is applied to enhance the feature maps dynamically. The scalar summary $s_c$, derived through GAP, represents the global activation strength of each channel, effectively capturing the overall significance of each channel's features across the entire spatial domain. By feeding this scalar summary into two fully connected layers with non-linear activation functions (ReLU) [29,30], the SE block learns the relative importance of each channel. The weight values $W_1$ and $W_2$, being learnable parameters, adapt dynamically during training to assign higher importance to channels that contribute significantly to detecting thermal anomalies, such as hotspots or defects, while suppressing less relevant features. This recalibration mechanism ensures that the feature maps are optimally weighted, enhancing the model's ability to focus on subtle patterns and anomalies often present in thermal imagery.

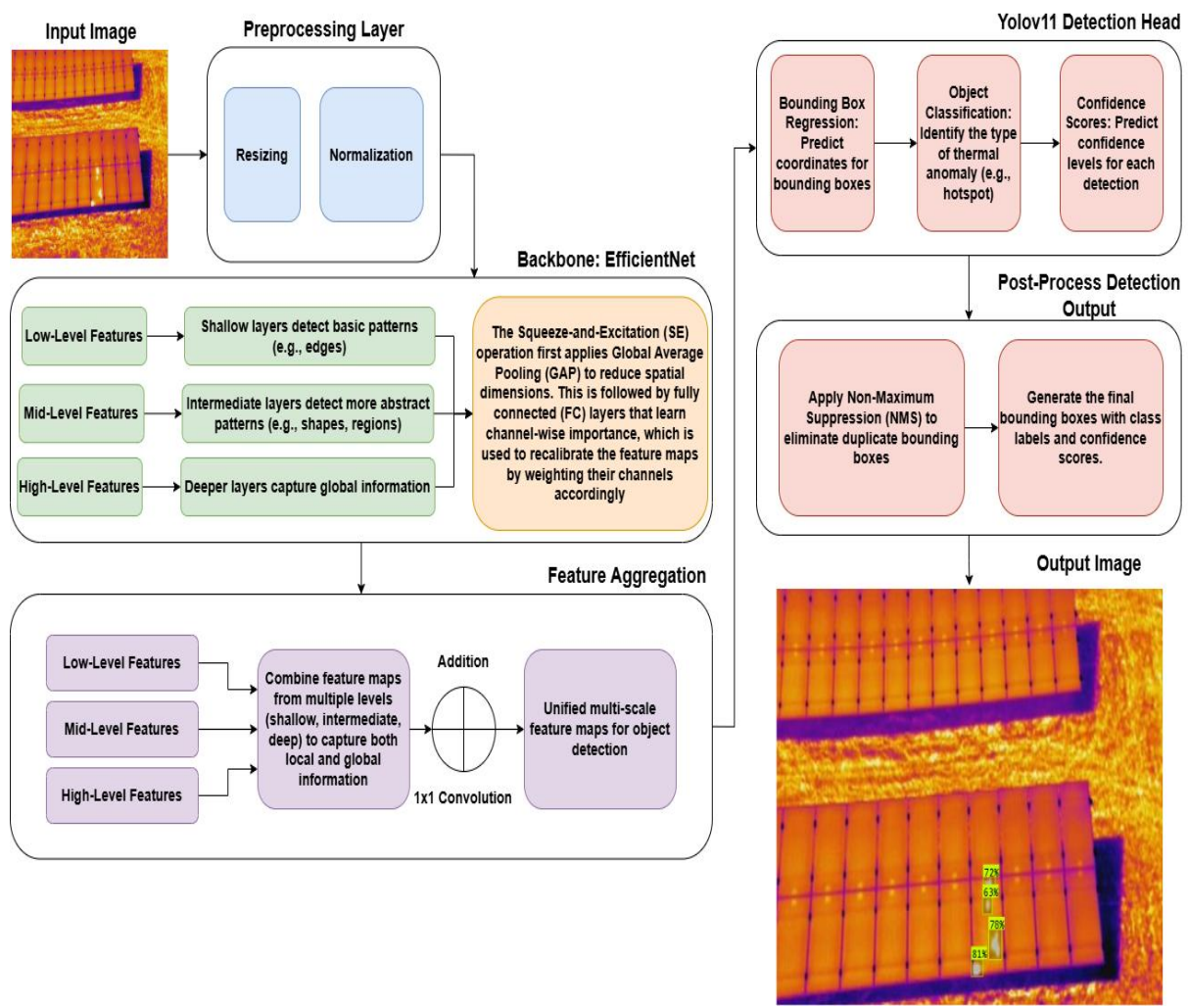

**Figure 2.** Schematic diagram of HOTSPOT-YOLO.

Feature maps from low, medium, and high layers are aggregated in the Feature Aggregation Block. These multi-scale features are fused using element-wise addition and dimensionality reduction through 1x1 convolutions and can be expressed as (4), where $F_{agg}$ is the unified multi-scale feature map of the low-level features ($F_{low-level}$), mid-level features ($F_{mid-level}$), and the high-level features ($F_{high-level}$), that all were detected from the input thermal PV image.

$$F_{agg} = Conv_{1\times1}(F_{low-level} + F_{mid-level} + F_{high-level}) \quad (4)$$

The feature aggregation process, as described in Equation (4), serves as a critical component in the HOTSPOT-YOLO architecture, ensuring the integration of hierarchical information from different feature levels. Low-level features ($F_{low-level}$) capture fine-grained spatial details such as edges and textures, which are essential for detecting small thermal anomalies. Mid-level features ($F_{mid-level}$) represent more abstract patterns, such as geometric shapes or regions of interest, while high-level features ($F_{high-level}$) capture broader contextual information that is crucial for recognizing larger or more complex structures. By combining these features using element-wise addition, the Feature Aggregation Block preserves the unique contributions of each level, ensuring that the resulting multi-scale representation leverages both local and global information. The inclusion of a 1x1 convolution further refines this process by performing dimensionality reduction, which not only unifies the feature map into a consistent representation ($F_{agg}$) but also minimizes computational overhead, making the model more efficient.

The YOLOv11 detection head processes the aggregated features to predict bounding boxes, class probabilities, and confidence scores. This head operates at three scales to detect small, medium, and large objects. For each grid cell in the feature map, the detection head predicts:

$$y = \{b, c, p\} \quad (5)$$

where $b = (x, y, w, h)$ are the bounding box coordinates, $c$ is the class label (thermal anomaly in the detected thermal PV image), and $p$ is the confidence score. The detection head minimizes the loss function as in (6), where $L_{box}$ measures the intersection over union (IoU) between predicted and ground truth boxes [31], $L_{class}$ is the classification loss, and $L_{conf}$ penalizes incorrect confidence predictions. The IoU is a standard metric used in object detection to evaluate how well the predicted bounding box overlaps with the ground truth bounding box.

$$L = L_{box} + L_{class} + L_{conf} \quad (6)$$

The final predictions are post-processed using NMS to eliminate overlapping bounding boxes and retain the most confident predictions using (7), where $B_{final}$ is the set of predicted bounding boxes. The NMS is a post-processing technique used in object detection to eliminate redundant bounding boxes that overlap significantly and retain only the most confident ones. It ensures that only the best bounding box for each object is kept, improving detection precision. As an example, the resulting thermal anomalies are presented as annotated bounding boxes on the original input image, as shown in Figure 2.

$$B_{final} = NMS(B, IoU\ threshold) \quad (7)$$

All parameters used in the HOTSPOT-YOLO model, including kernel sizes, strides, input/output dimensions, number of filters, and learnable parameters for each layer, are comprehensively detailed in Table 1. This table serves as a reference for the architectural components and their corresponding configurations, ensuring the model's reproducibility and transparency.

**Table 1.** Detailed layer-wise architecture of the HOTSPOT-YOLO model.

| Layer Name | Type | Kernel Size | Stride | Input Dimensions | Output Dimensions | Number of Filters | Parameters (M) | Description |
|---|---|---|---|---|---|---|---|---|
| **Input Image** | Thermal Image | - | - | 640x640x3 | 640x640x3 | - | - | Input thermal image captured by the drone. |
| **Preprocessing** | Resizing + Normalization | - | - | 640x640x3 | 224x224x3 | - | - | Image resized and normalized for EfficientNet. |
| **Conv1** | Conv2D | 3x3 | 2 | 224x224x3 | 112x112x32 | 32 | 0.9 | Initial convolution layer for feature extraction. |
| **Block1** | Depthwise Conv | 3x3 | 1 | 112x112x32 | 112x112x32 | 32 | 1.5 | EfficientNet bottleneck block with Depthwise Convolution. |
|  | Pointwise Conv | 1x1 | 1 | 112x112x32 | 112x112x64 | 64 | 4.5 | Combines depthwise feature maps. |
|  | SE Block | - | - | 112x112x64 | 112x112x64 | - | - | Squeeze-and-Excitation (SE) recalibrates feature maps. |
| **Block2** | Depthwise Conv | 3x3 | 2 | 112x112x64 | 56x56x64 | 64 | 1.2 | Down samples features while extracting spatial details. |
|  | Pointwise Conv | 1x1 | 1 | 56x56x64 | 56x56x128 | 128 | 5.6 | Combines depthwise feature maps. |
|  | SE Block | - | - | 56x56x128 | 56x56x128 | - | - | SE block for recalibration. |
| **Block3** | Depthwise Conv | 3x3 | 2 | 56x56x128 | 28x28x128 | 128 | 2.1 | Extracts high-level features while reducing spatial resolution. |
|  | Pointwise Conv | 1x1 | 1 | 28x28x128 | 28x28x256 | 256 | 7.3 | Combines depthwise feature maps. |
|  | SE Block | - | - | 28x28x256 | 28x28x256 | - | - | SE block for recalibration. |
| **Shallow Features** | Output from Block1 | - | - | 112x112x64 | 112x112x64 | - | - | Feature map from early layers (basic patterns). |
| **Intermediate Features** | Output from Block2 | - | - | 56x56x128 | 56x56x128 | - | - | Feature map capturing intermediate-level abstractions. |
| **Deep Features** | Output from Block3 | - | - | 28x28x256 | 28x28x256 | - | - | High-level feature map from deeper layers. |
| **Feature Aggregation** | Addition + 1x1 Conv | 1x1 | 1 | Multi-scale Inputs | Unified Map Size | 256 | 2.5 | Combines shallow, intermediate, and deep features into a unified multi-scale feature map. |
| **Detection Head (Small)** | Conv2D + Prediction | 3x3 | 1 | Unified Map Size | Prediction Size (Small) | 75 | 3.5 | Predicts bounding boxes, class labels, and confidence scores for small objects. |
| **Detection Head (Medium)** | Conv2D + Prediction | 3x3 | 1 | Unified Map Size | Prediction Size (Medium) | 75 | 3.5 | Predicts bounding boxes, class labels, and confidence scores for medium objects. |
| **Detection Head (Large)** | Conv2D + Prediction | 3x3 | 1 | Unified Map Size | Prediction Size (Large) | 75 | 3.5 | Predicts bounding boxes, class labels, and confidence scores for large objects. |
| **Post-Processing** | Non-Maximum Suppression | - | - | Prediction Outputs | Final Outputs | - | - | Eliminates duplicate predictions and finalizes bounding boxes with class labels and confidence scores. |
| **Output Image** | Annotated Image | - | - | Thermal Image + Boxes | Annotated Thermal Image | - | - | Displays thermal image with bounding boxes, class labels, and confidence scores for detected anomalies. |

## 2.3 HOTSPOT-YOLO Training Procedure and Dataset Description

To optimize the performance of the HOTSPOT-YOLO model, a rigorous training procedure was employed using publicly available thermal imagery datasets sourced from Roboflow. These datasets were specifically curated for thermal anomaly detection in solar PV systems, and all image annotations (bounding boxes and class labels) were performed using the annotation tools provided by Roboflow. The training process leveraged a combination of state-of-the-art data augmentation techniques, advanced optimization strategies, and careful parameter tuning to ensure robust model performance, as detailed in Table 2. The parameter tuning process involved systematic experimentation with key hyperparameters, including the learning rate, batch size, number of epochs, and the momentum of the optimizer.The dataset consists of 6000 images for training, 720 images for validation, and 450 images for testing, all of which were normalized and resized to $640 \times 640$ pixels to ensure consistency with the input requirements of the YOLOv11 framework. The annotations encompass thermal hotspots and other anomalies, representing regions of interest within the solar PV images. By employing horizontal flipping and random cropping during training, the diversity of the dataset was significantly enhanced, which is critical for improving the model's generalization capabilities.

Table 2. Training parameters for HOTSPOT-YOLO.

| Parameter | Value | Description |
|---|---|---|
| Optimizer | Adam | Optimization algorithm used to minimize the loss function. |
| Learning Rate | 0.001 | Initial learning rate for the optimizer. |
| Learning Rate Scheduler | Cosine Decay | Scheduler used to decay the learning rate over training epochs. |
| Batch Size | 16 | Number of samples processed per training step. |
| Number of Epochs | 200 | Total number of iterations over the entire training dataset. |
| Loss Function | YOLOv11 Loss | Combines classification, localization, and confidence losses for training the detection head. |
| Momentum | 0.9 | Momentum term used to accelerate gradient descent. |
| Weight Decay | 0.0005 | Regularization term to prevent overfitting by penalizing large weights. |
| Training Dataset Size | 3600 images | Total number of training images used for training the model. |
| Validation Dataset Size | 470 images | Total number of validation images used for model evaluation. |
| Image Augmentation | Horizontal Flip, Random Crop | Techniques applied to enhance dataset diversity and improve generalization. |
| Input Image Resolution | 640x640 pixels | Resolution to which all input images are resized before training. |
| Anchor Box Sizes | Auto-anchor (learned from data) | Auto-anchor box sizes for bounding box predictions, optimized form the dataset. |
| GPU Configuration | NVIDIA A100 GPU | GPUs used for training, specifying type and quantity. |
| Training Time | 2 hours | Approximate total time required for training the model on the specified hardware. |
| Evaluation Metric | $mAP$ (mean Average Precision) | Metric used to evaluate the model's performance during validation. |
| Early Stopping | Disabled | Training continues for all epochs unless overfitting is evident. |
| Gradient Clipping | Not used | Gradient clipping is optional for YOLOv11 and typically not required. |

An Adam optimizer [32,33] was used to minimize this loss with an initial learning rate of 0.001, which decays over epochs following a cosine decay scheduler to stabilize convergence. A batch size of 16 was utilized, leveraging the parallel processing capabilities of an NVIDIA A100 GPU on Google Colab to efficiently train the model over 200 epochs, as detailed in Table 2. The model's performance was evaluated using the mean Average Precision ($mAP$) metric, a standard evaluation method for object detection models. The $mAP$ is computed as (8). Appendix A presents the distribution and positional data of detected anomalies in the dataset, illustrating the spatial characteristics of bounding boxes and object dimensions for the thermal defect class.

$$mAP = \frac{1}{N}\sum_{i=1}^{N} AP_i \qquad (8)$$

where $N$ is the total number of object classes, and $AP_i$ represents the average precision for the $i-th$ class. The $mAP$ progression during training is illustrated in Figure 3, where the model achieved a final $mAP$ of 90.8% after 120 epochs (using the validation dataset). The learning curve demonstrates a steady improvement, with saturation occurring at approximately epoch 125, followed by small fluctuations, indicating robust training and convergence. Appendix B and Appendix C present two distinct sets of results showcasing the testing and validation performance of the model.

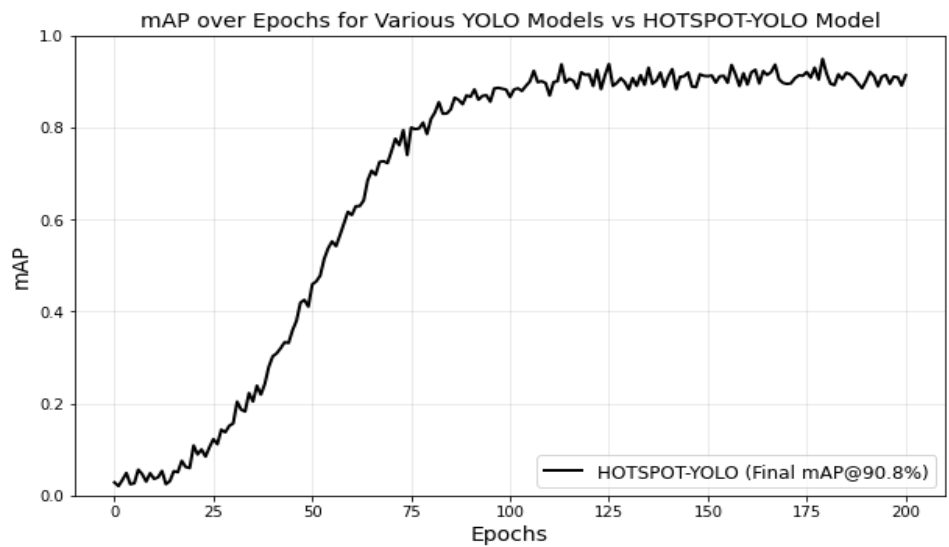

Figure 3. $mAP$ progression over 200 epochs for the HOTSPOT-YOLO model.

# 3. Results

## 3.1 Multi-Module Hotspot Detection (Scenario 1: Single Hotspot)

The HOTSPOT-YOLO model was tested on thermal images of PV arrays containing multiple modules, showcasing its robustness in detecting and localizing hotspots in complex scenarios. The results, as presented in Figure 4, demonstrate the model's ability to identify thermal anomalies with high confidence.

In Figure 4(a), the input image (left) displays a PV array with a thermal anomaly located in one of the central modules. The output image (right) shows that the model successfully identifies the hotspot, assigning it a confidence score of 87%, with precise localization. This example illustrates the model's ability to detect isolated thermal anomalies even within multi-module setups. Similarly, Figure 4(b) highlights a thermal image of another multi-module PV array. The model effectively detects a prominent hotspot with a confidence score of 90%, as shown in the output image (right). The accurate detection and localization of this anomaly, as compared to the input image (left), further underscore the model's reliability in real-world inspection scenarios. These results demonstrate the HOTSPOT-YOLO model's capability to detect hotspots across diverse PV module configurations, making it a reliable tool for large-scale thermal anomaly detection.

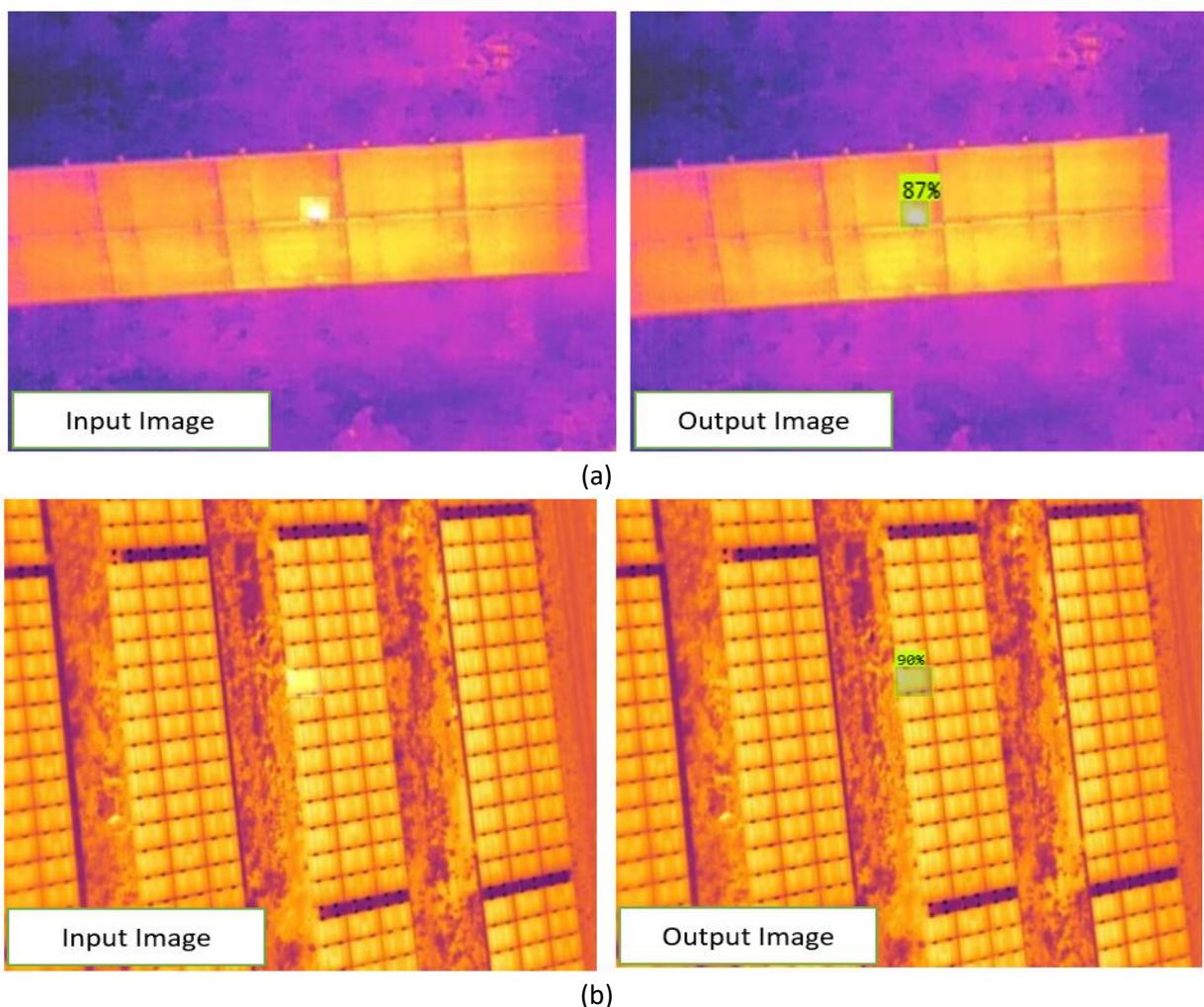

(a)

(b)

**Figure 4.** Multi-module hotspot detection results using the HOTSPOT-YOLO model. (a) Detection of a single hotspot in a PV array, with a confidence score of 87%. (b) Detection of a hotspot in a separate PV array, with a confidence score of 90%.

## 3.2 Multi-Module Hotspot Detection (Scenario 2: Multiple Hotspots)

The detection of multiple hotspots in PV arrays is a critical aspect of ensuring the operational efficiency and safety of large-scale solar installations. This subsection evaluates the performance of the HOTSPOT-YOLO model in identifying multiple hotspots within thermal images of multi-module PV arrays under varying environmental and thermal conditions.

The thermal image in Figure 5 is captured under very high irradiance conditions, evident from the uniform hot surface observed across all PV modules in the array. This high irradiance condition creates a baseline thermal pattern across the modules, making it challenging to distinguish true hotspots from regular thermal variations. Despite this, the HOTSPOT-YOLO model accurately detected four modules with significantly higher thermal anomalies, which deviate from the uniform pattern. The detected modules were identified with confidence scores ranging from 91% to 94%, showcasing the model's ability to focus on critical regions of interest. This result highlights the capability of HOTSPOT-YOLO to adapt to high-noise environments and isolate true hotspots amidst a uniformly heated background. The model's robustness in identifying non-patterned anomalies ensures its reliability in real-world conditions where environmental factors like solar irradiance may mask the visibility of faults.

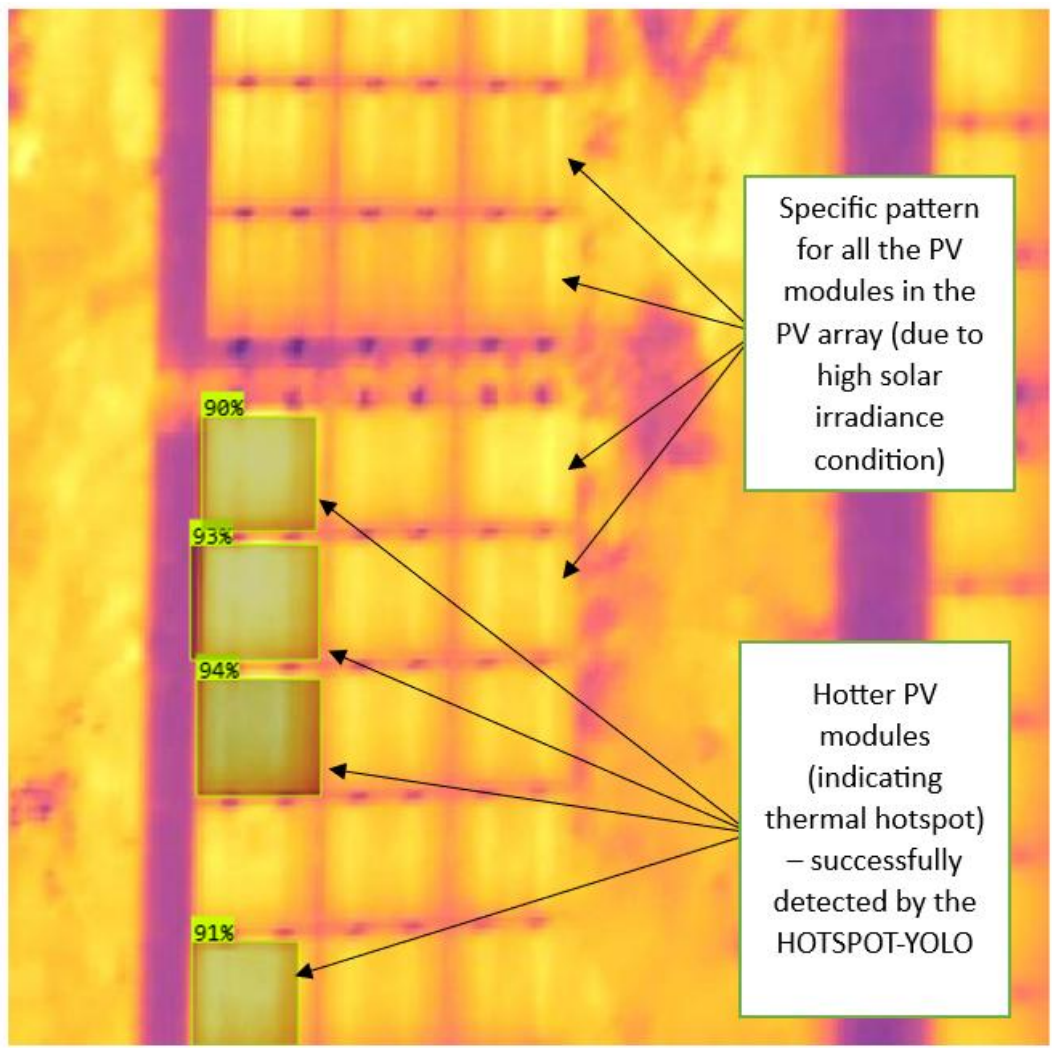

**Figure 5.** Detection of thermal hotspots in a multi-module PV array using HOTSPOT-YOLO. The model identifies localized hotspots with confidence scores ranging from 91% to 94%, while uniform thermal patterns caused by high solar irradiance conditions are observed across the array.

In Figure 6, the thermal image prominently displays a temperature scale bar on the right and a visually striking hot surface area below the PV array. While these features contribute to the overall thermal profile of the image, they do not represent true thermal anomalies within the PV modules themselves. The HOTSPOT-YOLO model demonstrated exceptional robustness by ignoring these irrelevant regions, focusing exclusively on the detection of true hotspots within the PV array. The identified hotspots, confined to two modules, were detected with confidence scores ranging from 78% to 79%, illustrating the model's precision in isolating genuine anomalies.

This capability to differentiate between thermal anomalies within the PV modules and irrelevant features, such as environmental heat sources or the thermal imaging camera's scale bar, is a critical advantage for automated inspection systems. It underscores the significance of the architectural enhancements in HOTSPOT-YOLO, particularly the integration of Squeeze-and-Excitation (SE) attention mechanisms and multi-scale feature aggregation. These mechanisms enable the model to prioritize spatially and semantically significant regions within the input images while suppressing irrelevant or redundant information.

The results from Figure 6 also highlight the model's ability to generalize effectively across diverse thermal conditions, including complex images with mixed thermal artifacts. By leveraging SE mechanisms, the model dynamically recalibrates feature importance based on the thermal context, ensuring that only the most relevant regions contribute to the final predictions. Moreover, the incorporation of multi-scale feature aggregation ensures that the model captures both fine-grained spatial details and broader contextual information, allowing it to adapt to varying thermal patterns. This fine-tuned focus makes HOTSPOT-YOLO a robust solution for real-world applications, where thermal noise and irrelevant patterns can often obscure meaningful anomalies.

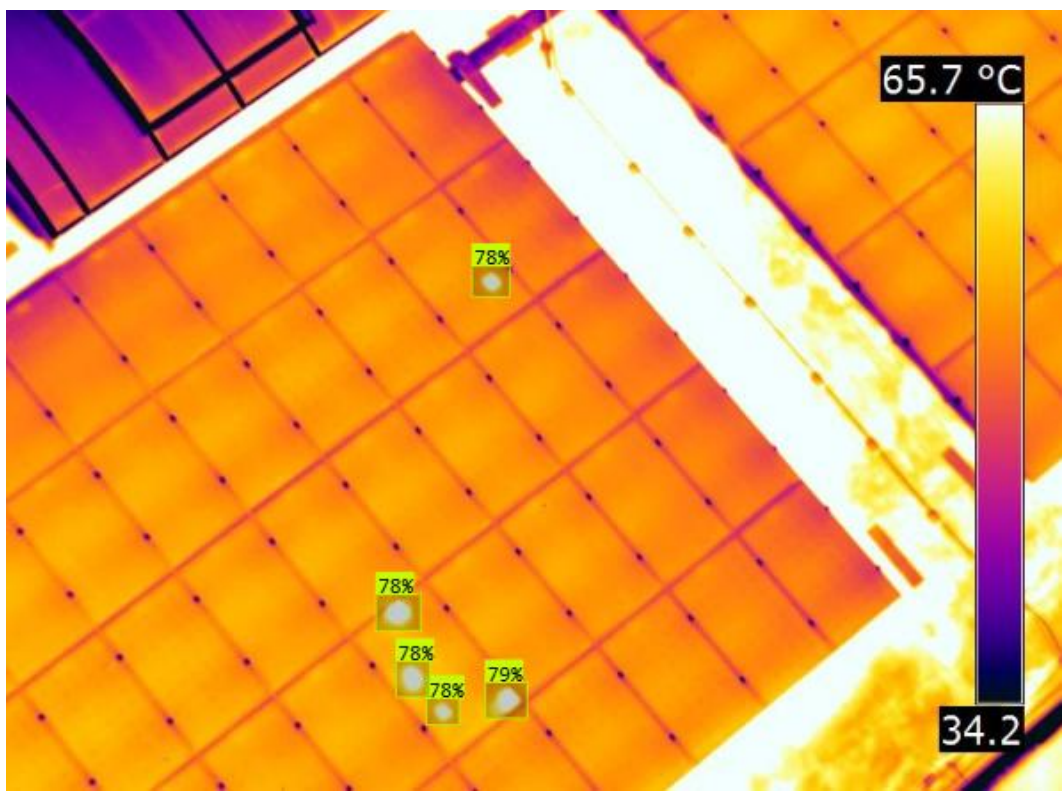

**Figure 6.** Detection of multiple thermal hotspots in a multi-module PV array using HOTSPOT-YOLO. The model identifies several hotspots with confidence scores ranging from 78% to 79%.

## 3.3 Multi-Module Hotspot Detection (Scenario 3: Robustness to Image Variations)

Ensuring robustness to image variations such as changes in brightness, contrast, and colour schemes is essential for the real-world deployment of object detection models. The HOTSPOT-YOLO model was evaluated under these challenging scenarios, and its performance is illustrated in Figure 7, which demonstrates its capability to maintain consistent hotspot detection across various transformations applied to the original image. Using a colour scheme instead of numeric temperature readings offers specific advantages, such as reducing the number of input channels by one-third (as colour images typically use three channels compared to single-channel numeric data), potentially improving computational efficiency without sacrificing the model's ability to identify thermal anomalies. Additionally, colour schemes can enhance visual interpretability for human operators, making it easier to identify patterns and anomalies during manual inspection.

The original thermal image- (Figure 7(a)) displays the PV array in its default state, with two hotspots detected. The detected hotspots are localized accurately, with confidence scores of 74% and 70%, respectively. This serves as the baseline for evaluating the model's performance under transformations. When the brightness and contrast of the image were reduced by 40%, as shown in Figure 7(b), the overall visibility of the PV modules diminished, simulating a low-light or partially obscured imaging condition. Despite this significant alteration, the HOTSPOT-YOLO model successfully detected both hotspots with confidence scores of 73%, only slightly reduced compared to the original detection.

The image was further transformed into grayscale as shown in Figure 7(c), removing all colour information, which is a vital cue in many object detection tasks. Even under these conditions, HOTSPOT-YOLO detected the same hotspots with slightly lower confidence scores of 68% and 67%. This minor reduction in detection accuracy indicates that the model can effectively utilize texture, structure, and spatial features extracted from grayscale images, rather than relying solely on colour-based information. Such robustness is critical for deployments in scenarios where colour information may be degraded or unavailable due to imaging constraints. Appendix D and Appendix E show additional examples of HOTSPOT-YOLO predictions using different image variations.

Additionally, real-world conditions, such as drone movement, camera vibrations, or environmental factors, often result in blurred thermal images with reduced edge clarity and lower contrast. Testing the HOTSPOT-YOLO model under such challenging conditions is critical to evaluate its reliability in detecting thermal anomalies. Figure 8(a) and Figure 8(b) illustrate the model's performance when subjected to image blurring, showcasing its robustness in identifying hotspots in degraded visual environments.

In the example shown in Figure 8(a), the image exhibits significant blurring, with unclear module edges and a substantial reduction in overall contrast. Despite these challenges, the HOTSPOT-YOLO model successfully detected multiple hotspots distributed across different modules, with confidence scores ranging from 53% to 71%. The decreased confidence scores, compared to detections in clear images, reflect the impact of reduced edge clarity and low contrast on the model's ability to localize and classify hotspots. However, the successful detection of all relevant hotspots demonstrates the robustness of the feature extraction pipeline, particularly the EfficientNet backbone and attention mechanisms, which focus on semantically relevant features rather than solely relying on edge details.

Similarly, Figure 8(b) presents a less severe blurring effect compared to Figure 8(a), but still shows reduced edge sharpness and contrast. The model accurately detected three hotspots across the PV modules, achieving confidence scores of 79%, 87%, and 88%. The higher confidence scores in this case, as compared to Figure 8(a), indicate that the model's detection accuracy improves with even a slight increase in edge clarity. This highlights the sensitivity of the detection process to image quality, while also confirming the model's adaptability to varying degrees of visual degradation.

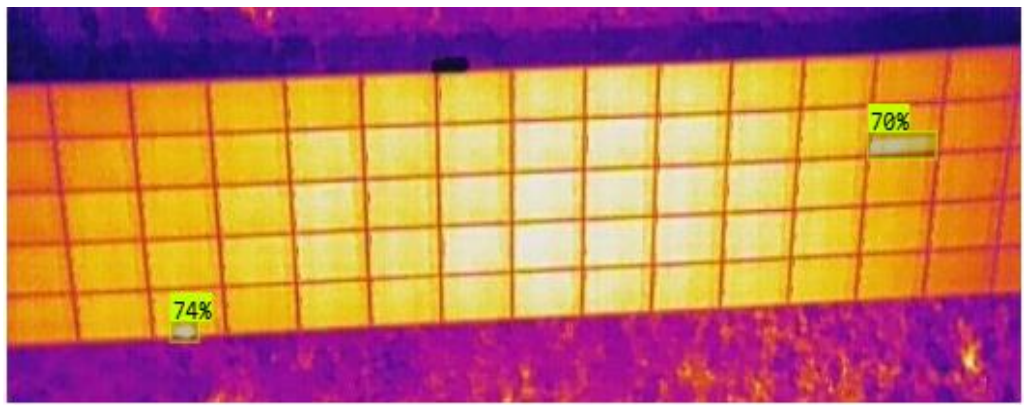

(a)

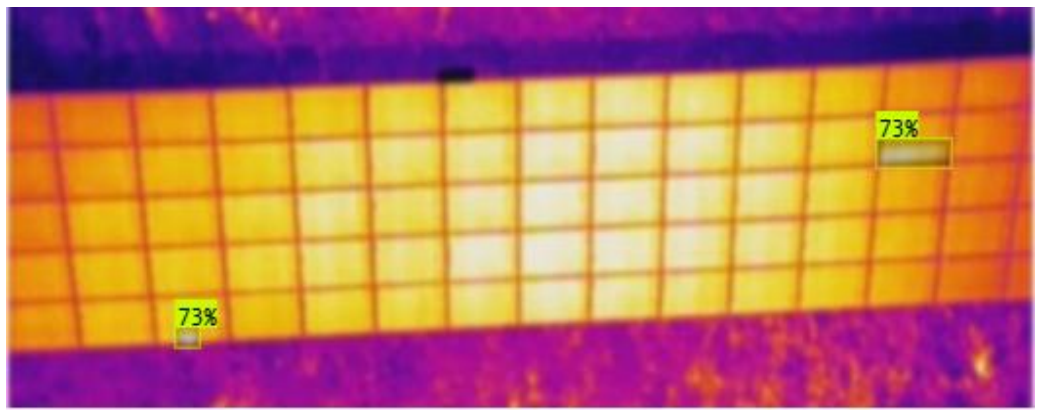

(b)

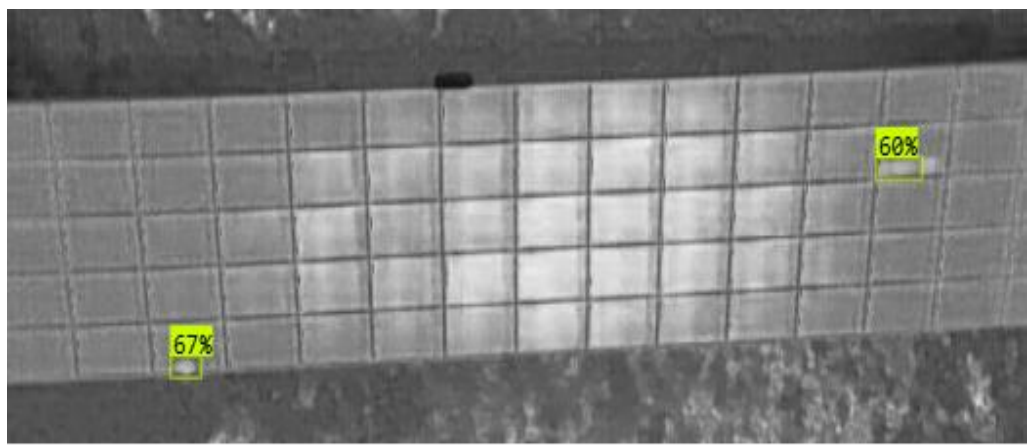

(c)

**Figure 7.** Evaluation of HOTSPOT-YOLO under varying image transformations. (a) Original thermal image with two detected hotspots (confidence scores: 74% and 70%). (b) Image with -40% brightness and -40% contrast, where the detection accuracy remains consistent (confidence scores: 73%). (c) Grayscale transformation, showing a minor reduction in confidence scores (68% and 67%), demonstrating the model's robustness to colour and brightness variations.

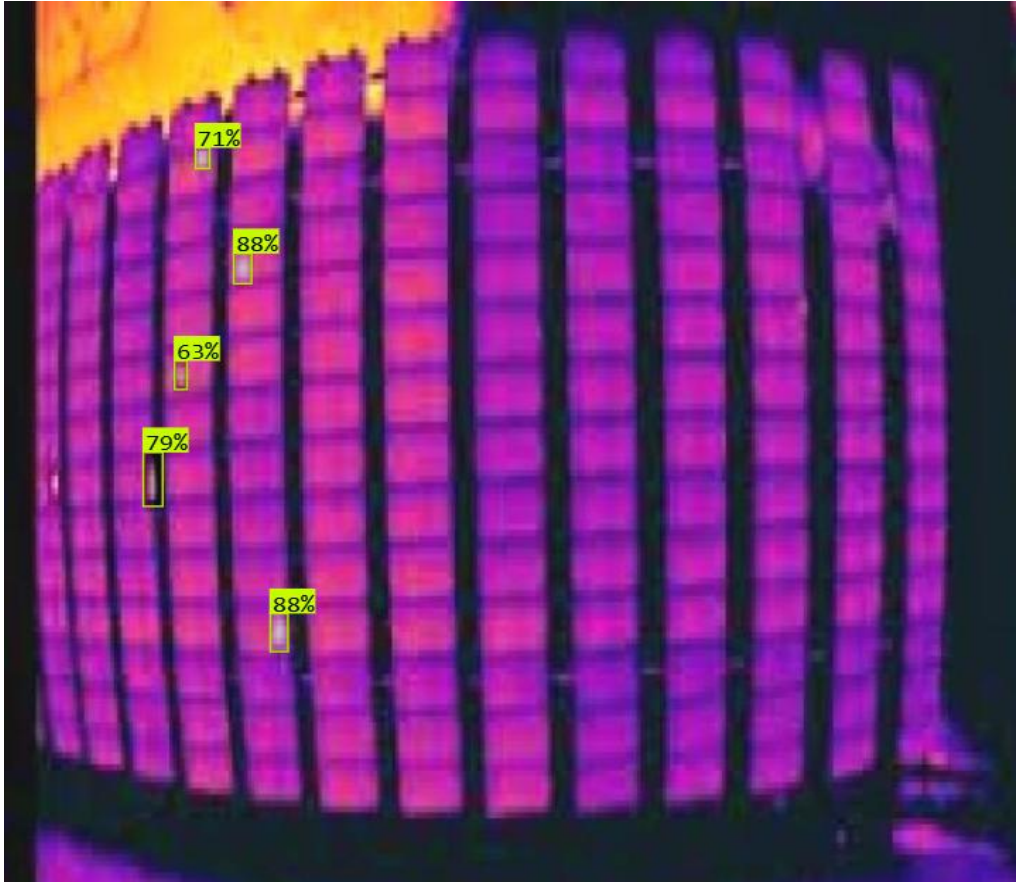

(a)

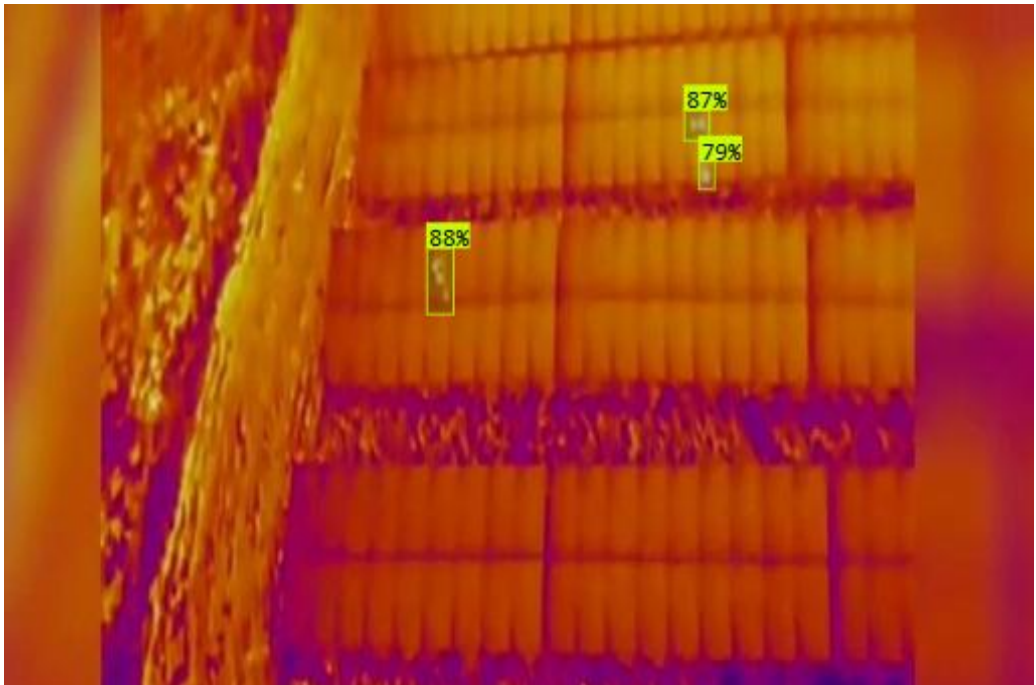

(b)

**Figure 8.** HOTSPOT-YOLO performance under image blurring conditions. (a) Detection in a highly blurred thermal image with low edge clarity and reduced contrast, identifying hotspots with confidence scores ranging from 53% to 71%. (b) Detection in a moderately blurred image, successfully identifying three hotspots with confidence scores of 79%, 87%, and 88%, showcasing the model's robustness to varying levels of visual degradation.

## 3.4 Multi-Module Hotspot Detection (Scenario 4: Ground-Based Thermal Imaging)

While the HOTSPOT-YOLO model was solely trained and validated using drone-captured thermal images, its robustness was further evaluated using thermal images taken from stationary, ground-based thermal cameras. This test scenario, depicted in Figure 9(a) and Figure 9(b), demonstrates the model's capability to detect hotspots under different imaging conditions, such as non-uniform angles, varied focus, and imperfect image orientation. These conditions mimic real-world challenges that may arise during manual ground-based inspections of PV arrays.

In Figure 9(a), the thermal image captures multiple PV modules from a ground-level perspective, introducing challenges such as uneven angles and partial occlusions. Despite these complexities, HOTSPOT-YOLO successfully detected the majority of the hotspots across the PV array, with confidence scores ranging from 63% to 82%. Notably, hotspots in modules near the center of the image were detected with higher accuracy due to their better alignment and focus. However, some hotspots, particularly at the edges of the image, remained undetected. This limitation can be attributed to the model's reliance on features learned from drone-captured images, where the angles and focus are more uniform and optimized for overhead views.

The close-up ground-based image in Figure 9(b) shows a smaller section of the PV array with varied focus and rotation. The model successfully detected two prominent hotspots with confidence scores of 83% and 85%, demonstrating its ability to adapt to localized thermal anomalies. However, a third hotspot was detected with a significantly lower confidence score of 28%, indicating the model's reduced accuracy for anomalies located at image edges or areas with uneven focus. The lower confidence score for the distant hotspot underscores the impact of ground-based imaging variations on the model's detection accuracy. This emphasizes the importance of aligning the input data characteristics with the model's training dataset to achieve optimal results. Some more results of ground-based thermal PV modules thermal detection are available in Appendix F.

Despite being trained exclusively on drone-captured images, HOTSPOT-YOLO demonstrated the ability to detect hotspots in ground-based thermal images. This highlights the versatility of the model's architecture, particularly its robust feature extraction capabilities. The significant variation in detection accuracy observed in Figures 9(a) and 9(b) reveals the sensitivity of the model to angle, orientation, and focus differences between the training data and test images. These factors must be carefully considered when deploying HOTSPOT-YOLO for ground-based inspections. Hotspots located at the edges of the image or in regions with oblique angles are more likely to remain undetected or be assigned lower confidence scores. This limitation arises from the inherent difference in the spatial and thermal patterns captured by drone and ground-based imaging systems. However, since the drone typically flies in a predefined pattern, it often captures multiple images of the same area from different perspectives, providing additional opportunities for detecting previously missed hotspots. This redundancy in coverage helps mitigate the risk of missing anomalies, improving the overall reliability of the detection process.

To enhance the performance of HOTSPOT-YOLO on ground-based images, additional fine-tuning using an expanded dataset that includes ground-based thermal images captured under varying conditions should be implemented. Such a dataset should encompass a wide range of scenarios, including diverse angles, orientations, focal depths, and lighting variations, to bridge the gap between the drone-based and ground-based imaging domains. Fine-tuning the model with these datasets would allow it to learn features specific to ground-based imagery, such as edge distortions, perspective shifts, and non-uniform thermal patterns. This adaptation would not only improve the detection accuracy for challenging scenarios but also increase the model's generalizability and robustness, making it a more versatile tool for comprehensive solar PV inspections across different deployment settings.

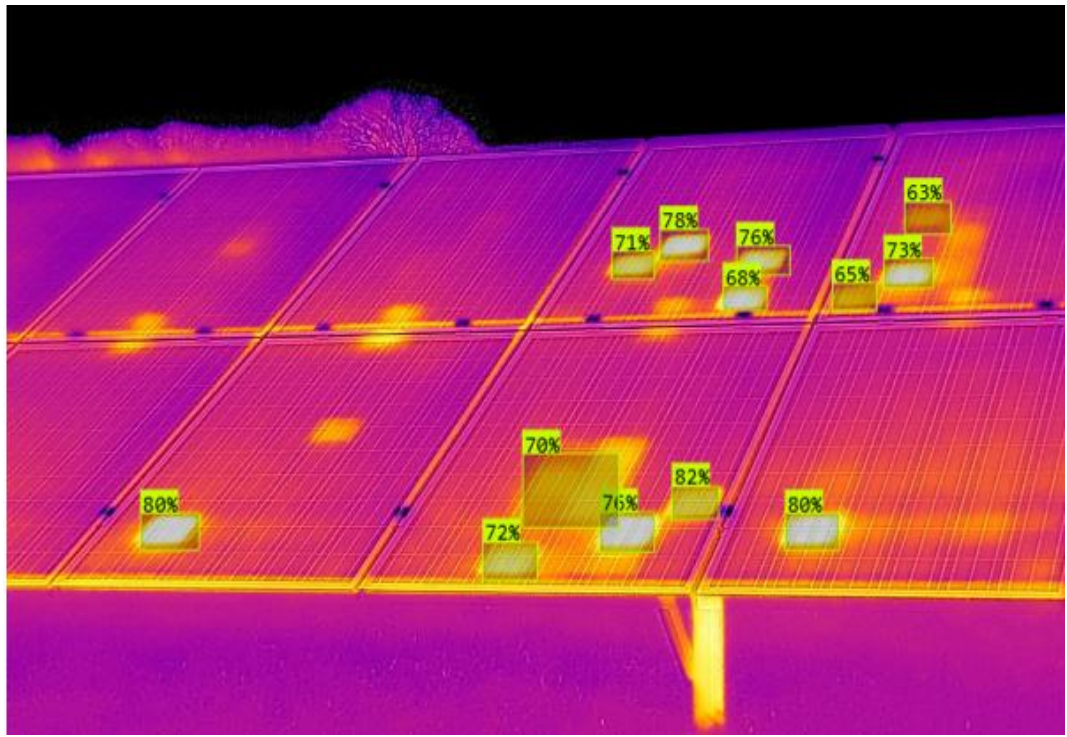

(a)

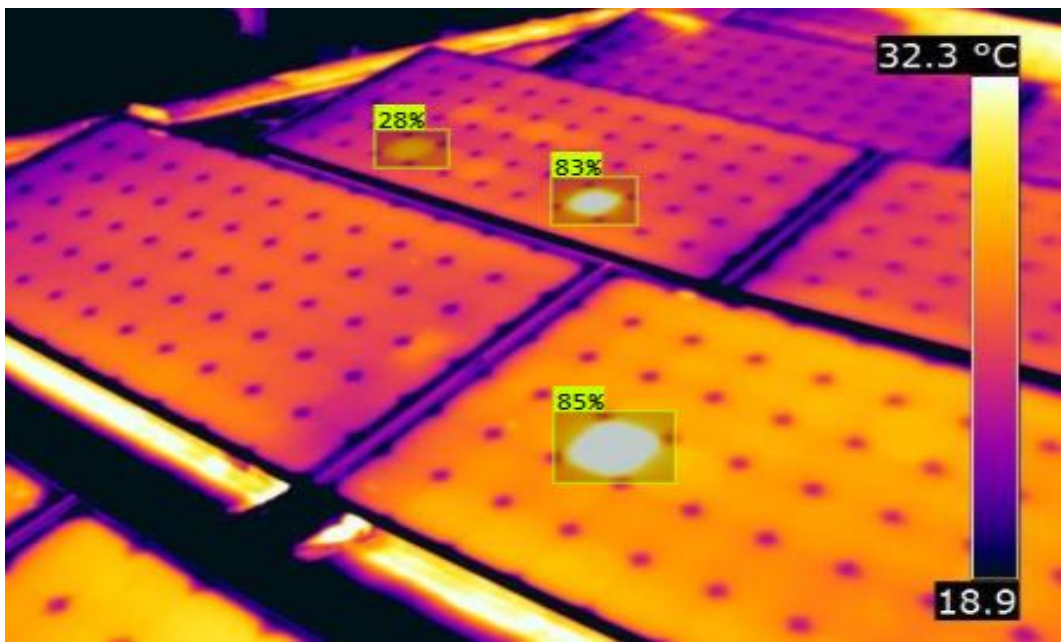

(b)

**Figure 9.** HOTSPOT-YOLO detection performance on ground-based thermal images. (a) Wide-angle thermal image with multiple detected hotspots (confidence scores: 63% to 82%), showing limitations in detecting edge anomalies due to angle and focus differences. (b) Close-up thermal image with rotation and focus variations, detecting two prominent hotspots (83% and 85%) and one distant hotspot with a lower confidence score of 28%, illustrating the impact of imaging variations on detection accuracy.

## 3.6 Benchmark Analysis

The proposed HOTSPOT-YOLO model was benchmarked against several state-of-the-art object detection algorithms to highlight its performance in detecting thermal anomalies in solar PV systems. These models, including Cascade RCNN [34], CenterNet [35], Faster RCNN [36], YOLOv5m [37], YOLOv9m [38], and YOLOv11m [39], are widely recognized for their effectiveness in object detection tasks across various domains. The benchmarking results, presented in Table 3, assess each model's detection accuracy ($mAP@0.5$), computational efficiency (FLOPs), model complexity (number of parameters), and inference speed (ms), providing a comprehensive comparative analysis.

The proposed HOTSPOT-YOLO achieves the highest detection accuracy with a $mAP$ of 90.8%, significantly outperforming YOLOv11m, which achieves 84.9% and is the next best-performing model. This notable improvement is attributed to architectural modifications, including the integration of EfficientNet as the backbone and Squeeze-and-Excitation attention mechanisms, which enhance the model's ability to focus on thermally significant regions like hotspots in solar modules. In contrast, traditional models such as Cascade RCNN and CenterNet exhibit lower accuracy, with $mAP$ scores of 74.2% and 68.5%, respectively, reflecting their limitations in handling small-scale, subtle thermal anomalies.

The $mAP$ progression for various YOLO models compared to the HOTSPOT-YOLO model is shown in Figure 10, illustrating a detailed comparison of their $mAP$ across 200 epochs. HOTSPOT-YOLO achieves the highest final $mAP$ at 90.8%, demonstrating a significant improvement over YOLOv11m (84.9%), YOLOv9m (80.3%), and YOLOv5m (81.4%). This result emphasizes the proposed model's capability to better capture critical features and detect subtle thermal anomalies in solar PV systems, making it a robust solution for anomaly detection tasks. The HOTSPOT-YOLO curve shows a smooth and consistent growth in $mAP$, reaching higher accuracy earlier during training and stabilizing without significant fluctuations after 125 epochs. In contrast, YOLOv9m exhibits more variability, indicating that it may struggle with noise or suboptimal feature extraction during training. YOLOv11m shows a stable curve but lags behind HOTSPOT-YOLO in final accuracy, underscoring the advantages introduced by HOTSPOT-YOLO's enhancements, such as the EfficientNet backbone and Squeeze-and-Excitation attention mechanisms.

Another notable observation from Figure 10 is the faster convergence rate of HOTSPOT-YOLO compared to other models. While YOLOv5m and YOLOv9m require more epochs to stabilize, HOTSPOT-YOLO quickly reaches its optimal detection accuracy, making it more suitable for real-time and resource-constrained scenarios. Additionally, the stability and reliability of HOTSPOT-YOLO across the training process highlight its robustness in handling diverse image variations and challenging features inherent in thermal imaging datasets.

**Table 3.** Comparison of HOTSPOT-YOLO with state-of-the-art object detection models

| Algorithms | $mAP$ (%) | Parameters (M) | FLOPs (G) | Time (ms) |
|---|---|---|---|---|
| Cascade RCNN [34] | 74.2 | 68.93 | 80.15 | 25.57 |
| CenterNet [35] | 68.5 | 28.89 | 142.13 | 27.62 |
| Faster RCNN [36] | 75.8 | 41.12 | 41.12 | 27.43 |
| YOLOv5m [37] | 81.4 | 18.12 | 15.83 | 22.35 |
| YOLOv9m [38] | 80.3 | 25.30 | 105.2 | 23.76 |
| YOLOv11m [39] | 84.9 | 38.22 | 29.61 | 29.61 |
| HOTSPOT-YOLO | 90.8 | 36.10 | 25.53 | 25.22 |

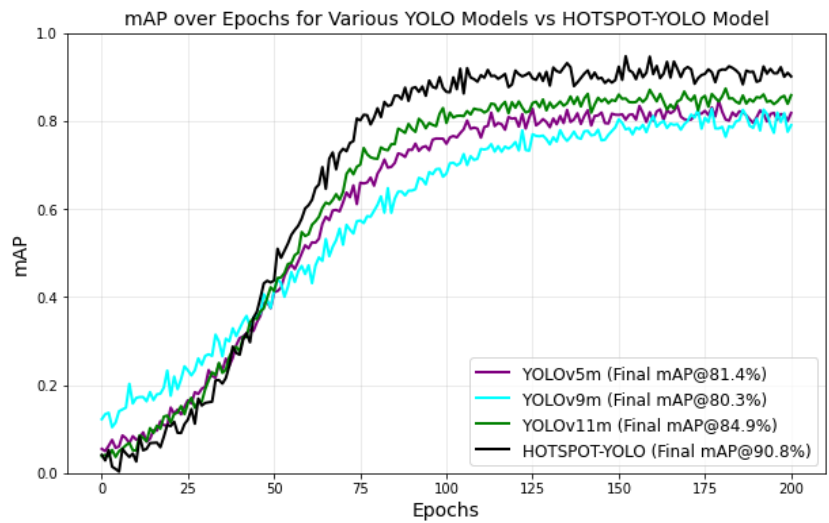

**Figure 10.** $mAP$ progression over epochs for various YOLO models vs. HOTSPOT-YOLO model. The figure highlights the superior performance of HOTSPOT-YOLO with the highest final $mAP@90.8\%$, demonstrating consistent improvements across training epochs compared to other YOLO variants.

From the perspective of model complexity, HOTSPOT-YOLO demonstrates an optimal balance with a parameter count of 36.10M, making it efficient while maintaining high accuracy. Compared to YOLOv11m, which has 38.22M parameters, HOTSPOT-YOLO achieves better accuracy with fewer parameters, showcasing the effectiveness of its architectural enhancements. On the other hand, models such as Cascade RCNN and Faster RCNN have significantly higher parameter counts, at 68.93M and 41.12M respectively, which lead to increased computational costs without commensurate gains in detection accuracy.

In terms of computational efficiency, the proposed HOTSPOT-YOLO model requires 25.53 GFLOPs, which is considerably lower than YOLOv9m (105.2 GFLOPs) and Cascade RCNN (80.15 GFLOPs). This computational lightness is crucial for real-time deployment scenarios, such as drone-based thermal imaging, where processing resources are often constrained. The integration of EfficientNet and feature aggregation mechanisms enables the model to maintain high performance while minimizing computational demands, further reinforcing its suitability for operational scalability.

The inference time of HOTSPOT-YOLO is 25.22 ms, placing it among the fastest models evaluated. While YOLOv5m achieves a slightly faster inference time of 22.35 ms, HOTSPOT-YOLO remains faster than YOLOv11m (29.61 ms) and RCNN-based models such as Cascade RCNN (25.57 ms) and Faster RCNN (27.43 ms). This speed ensures the model's applicability in real-time scenarios, such as drone-based PV inspections, where rapid image processing and anomaly detection are critical.

The results presented in this benchmarking analysis underscore the robustness and scalability of the HOTSPOT-YOLO model. Its superior detection accuracy, combined with its computational efficiency and lightweight design, makes it a compelling choice for detecting subtle thermal hotspots in PV arrays. The EfficientNet backbone reduces FLOPs while retaining strong feature extraction capabilities, and the SE attention mechanisms enhance the model's ability to focus on relevant regions. These features enable HOTSPOT-YOLO to adapt to multi-scale inputs and outperform competing models in both accuracy and efficiency. Furthermore, its real-time inference capability highlights its potential for deployment in large-scale PV inspections, providing a practical and effective solution for detecting thermal anomalies.

Table 4 provides a further comprehensive comparison of existing PV hotspot detection techniques, highlighting key methodologies, features, advantages, and limitations. These methods range from traditional approaches, such as IR imaging with partial shading analysis, to advanced solutions, including CNN and modified YOLO frameworks. For instance, [40] advanced real-time monitoring capabilities by employing a multi-task fusion approach with a modified YOLOv3 model. Researchers in [40], further utilized IR imaging combined with partial shading analysis to simulate and detect hotspots, offering a practical and cost-effective solution for specific scenarios. Similarly, [42] leveraged color image descriptors with infrared thermography to identify hotspots, demonstrating high accuracy for small datasets. The authors of [13] introduced a CNN-based method for large-scale condition monitoring, providing automated hotspot classification, albeit with significant computational overhead. The limitations observed across these methods, such as reliance on high-quality training data, scalability challenges, and computational intensity, underscore the need for efficient, robust solutions. In this context, our HOTSPOT-YOLO model, which integrates a lightweight YOLO framework, delivers superior performance with a high $mAP$ of 90.8% and scalable capabilities, addressing many of the constraints present in prior works.

Table 4. Comparative analysis of PV hotspots detection methods vs. HOTSPOT-YOLO

| Ref.; Year | Model | Key Features | Advantages | Limitations |
|---|---|---|---|---|
| [40]; 2023 | Modified YOLOv3 with Multi-Task Fusion. | Multi-task fusion for detecting hotspots in thermal IR images. | Improved accuracy and adaptability for real-time monitoring. | Limited field testing; requires high-quality training data. |
| [41]; 2021 | IR Imaging with Partial Shading Analysis. | Uses partial shading to simulate hotspots with IR imaging tools. | Practical for specific environmental scenarios; cost-effective. | Dependent on accurate shading patterns for results. |
| [42]; 2022 | Color Image Descriptors with IR Thermography. | Employs color-based descriptors to analyze thermal IR images. | High accuracy for small datasets; low computational cost. | Susceptible to noise in mixed data; limited scalability. |
| [13]; 2020 | CNN-Based Condition Monitoring. | Leverages convolutional neural networks for anomaly detection in IR data. | Automated large-scale anomaly detection and hotspot classification. | Requires significant computational resources for training. |
| This work; 2025 | HOTSPOT-YOLO (Enhanced YOLOv11). | Lightweight YOLO with EfficientNet backbone and SE attention mechanisms. | High $mAP$ (90.8%), low computational cost, scalable to large datasets. | Limited testing on ground-based thermal imaging datasets. |

## 4. Conclusion

This study addresses the critical need for efficient and accurate thermal anomaly detection in solar PV systems by introducing HOTSPOT-YOLO, an enhanced deep learning model optimized for drone-based inspections. The research demonstrates significant contributions to the field through architectural innovations, including the integration of an EfficientNet backbone for optimized feature extraction and SE attention mechanisms for precise focus on thermal anomalies. These enhancements enable the model to detect subtle defects, such as hotspots, with exceptional accuracy while maintaining computational efficiency.

Key findings reveal that HOTSPOT-YOLO achieves a $mAP$ of 90.8%, an improvement of 5.9% over the YOLOv11 baseline, alongside a reduction of 2.12 million parameters, making it computationally lightweight and practical for real-time deployment. Robustness tests highlight the model's adaptability to challenging conditions, including brightness variations, noise, and low-contrast images, confirming its utility for large-scale solar PV inspections. Benchmarking against leading object detection models further underscores HOTSPOT-YOLO's superior performance, achieving a notable balance between accuracy, speed, and resource efficiency. The implications of this work extend beyond immediate application in PV systems, offering insights into the potential of lightweight, attention-enhanced architectures for broader thermal imaging challenges. By effectively addressing scalability and resource constraints, HOTSPOT-YOLO represents a step forward in automating fault detection and enhancing the reliability of renewable energy technologies.

Building on the strengths of this study, future research will aim to extend HOTSPOT-YOLO's capabilities in the following areas: (1) Fine-tuning the model for ground-based thermal imagery and oblique-angle datasets to expand its applicability to diverse inspection scenarios; (2) Optimizing the model for integration with drone-edge processors to minimize latency and energy consumption during field inspections; (3) Leveraging detected anomalies to build predictive frameworks that enable proactive maintenance and reduce operational downtime; (4) Incorporating geographically varied datasets, including different environmental conditions and PV system configurations, to improve the model's robustness and scalability; and (5) Combining the detection framework with advanced AI-based diagnostic tools, such as digital twins, to support decision-making and long-term system optimization.

# Appendix A

This figure illustrates the spatial distribution and characteristics of detected thermal anomalies within the dataset:

**Top-left (Big Blue Square):** The blue square represents the total number of detected anomaly instances in the dataset. Its uniform coloration indicates that the dataset contains a consistent labeling of anomalies, which is crucial for training robust object detection models.

**Top-right (Bounding Box Overlap):** The overlay of bounding boxes visualises the positional distribution of anomalies. The clustering toward the center suggests that most anomalies are concentrated within the central region of the images, potentially influenced by the drone's typical flight path or image framing during data collection.

**Bottom-left (Scatter Plot of X and Y Coordinates):** This plot shows the normalized positional distribution of anomalies along the X and Y axes within the image space. The relatively uniform distribution highlights that anomalies are detected across the entire image, though slight clustering patterns may reflect operational constraints or common defect locations.

**Bottom-right (Height vs. Width Distribution):** This plot presents the normalized dimensions of bounding boxes (height vs. width). The dense cluster in the lower-left corner indicates that most anomalies are relatively small, reflecting the localized nature of thermal defects in solar PV systems.

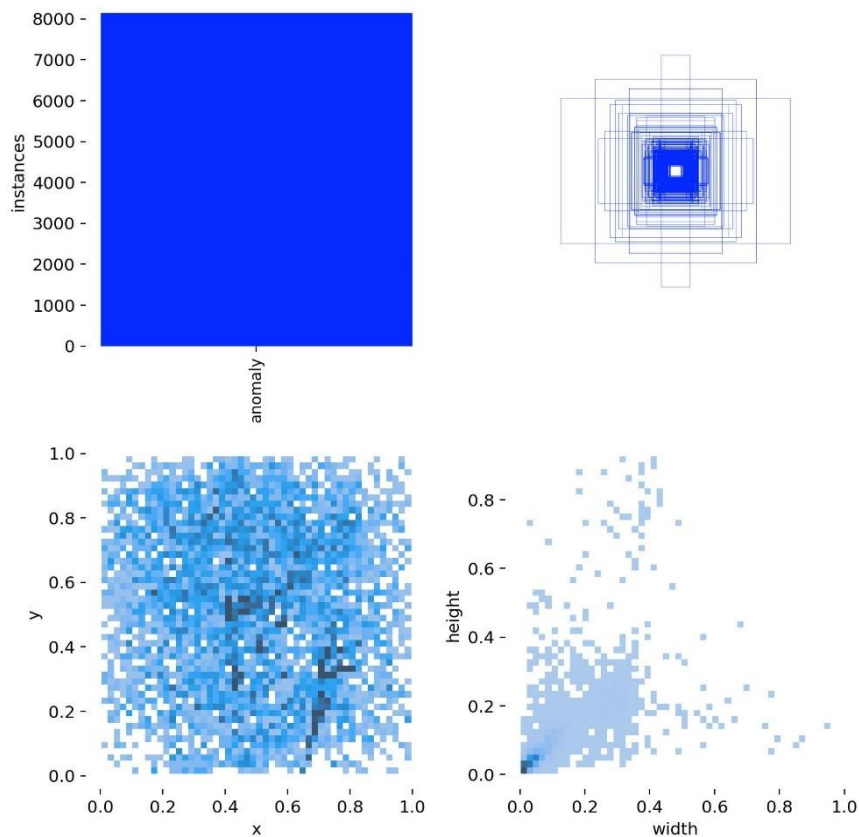

# Appendix B

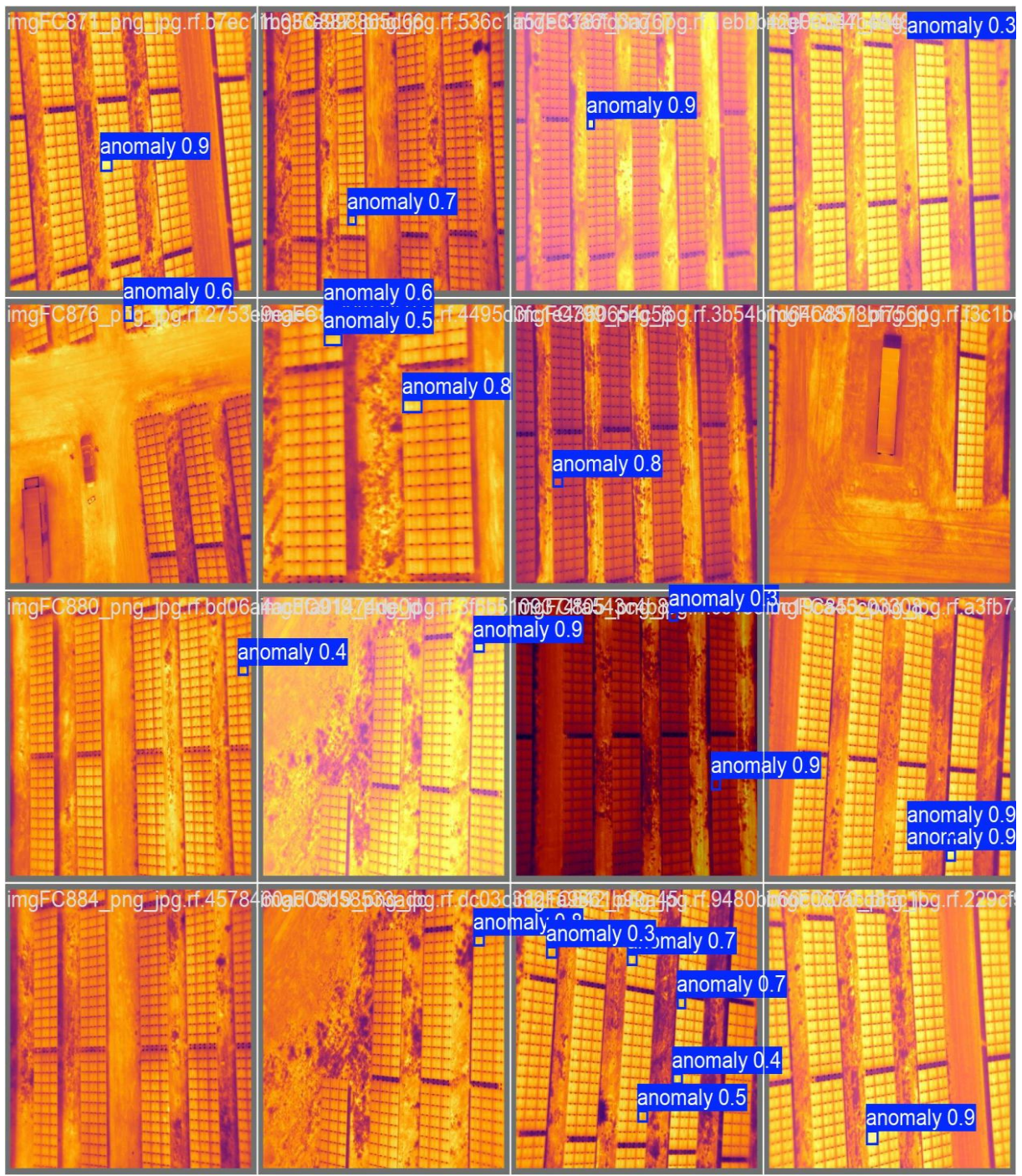

# Appendix C

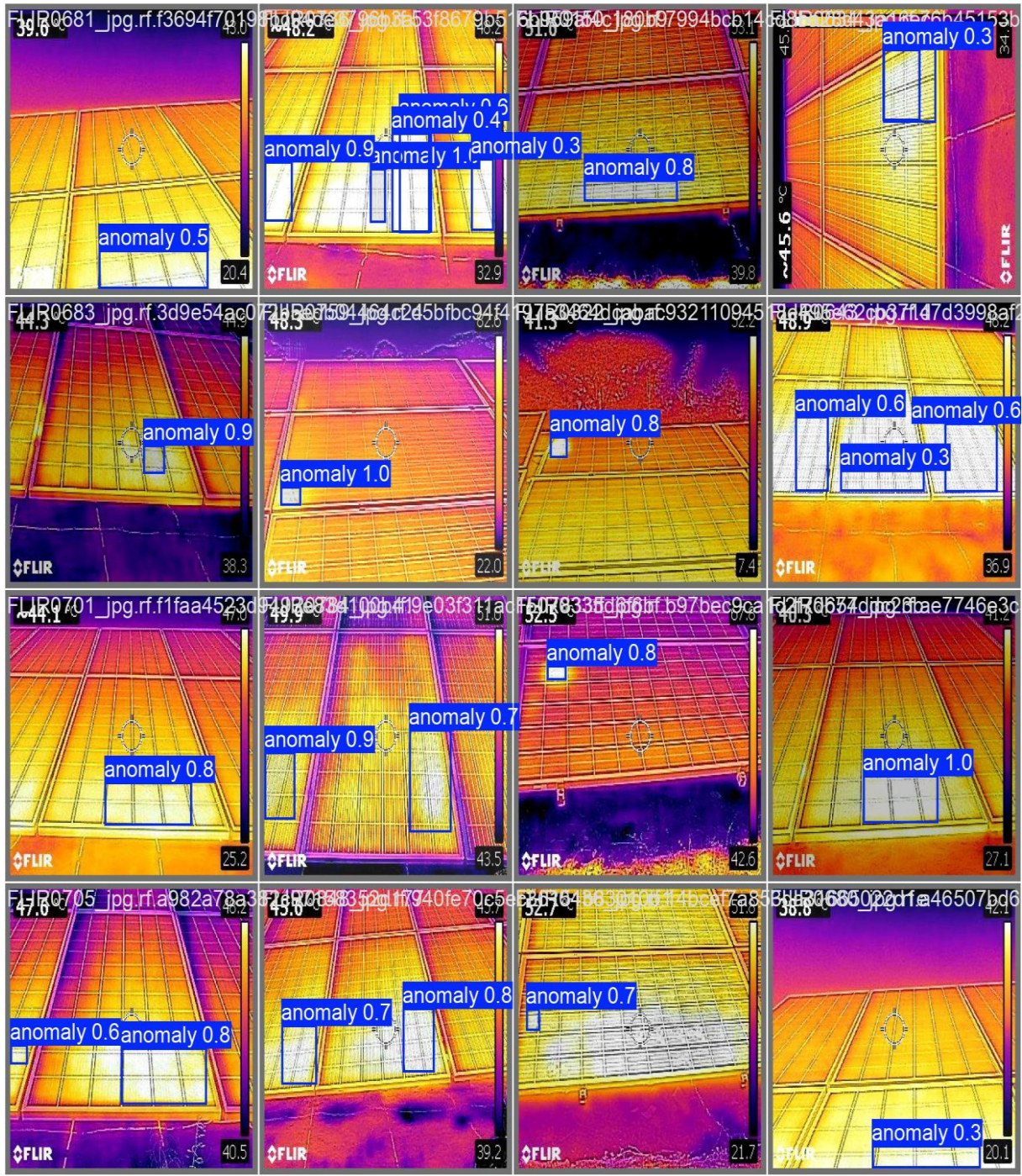

# Appendix D

| Brightness | Contrast | Output Image |
|---|---|---|
| **Original Image (Without Any Modifications)** | | 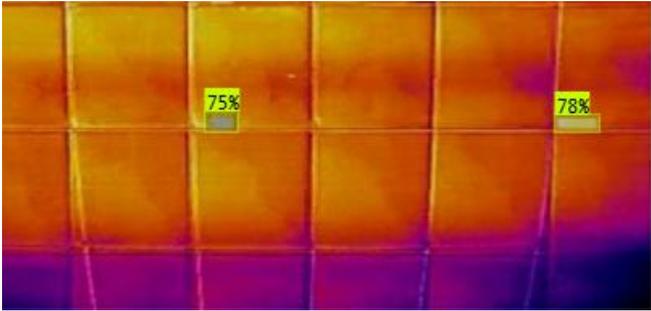 |
| **+25%** | **0%** | 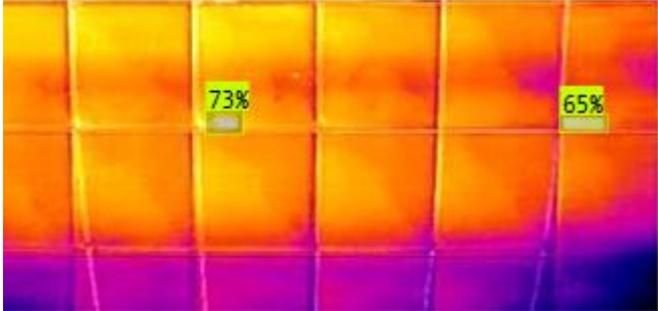 |
| **0%** | **+25%** | 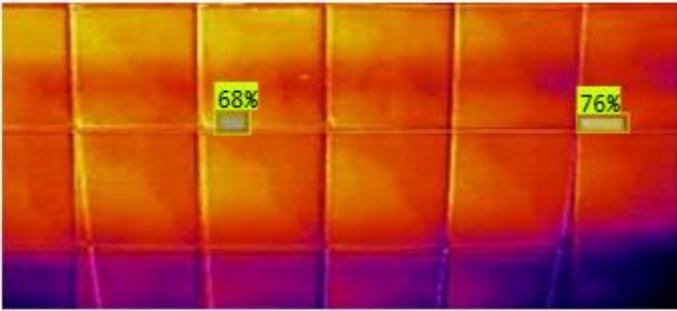 |
| **-35%** | **-35%** | 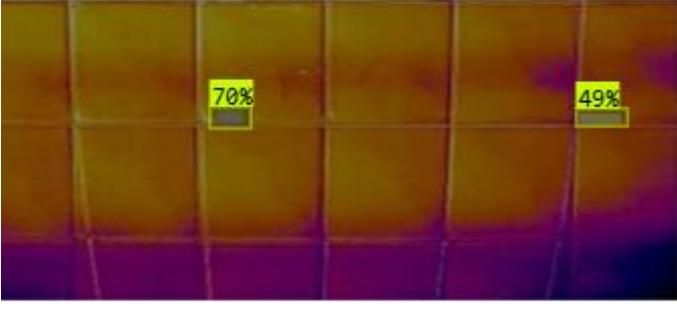 |
| **Original Image Convertered to Grayscale Iamge** | | 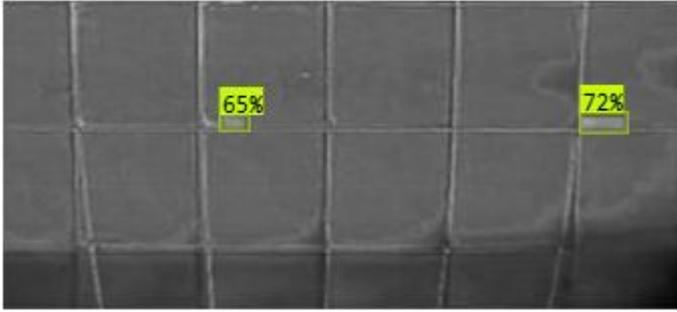 |

**Appendix E**

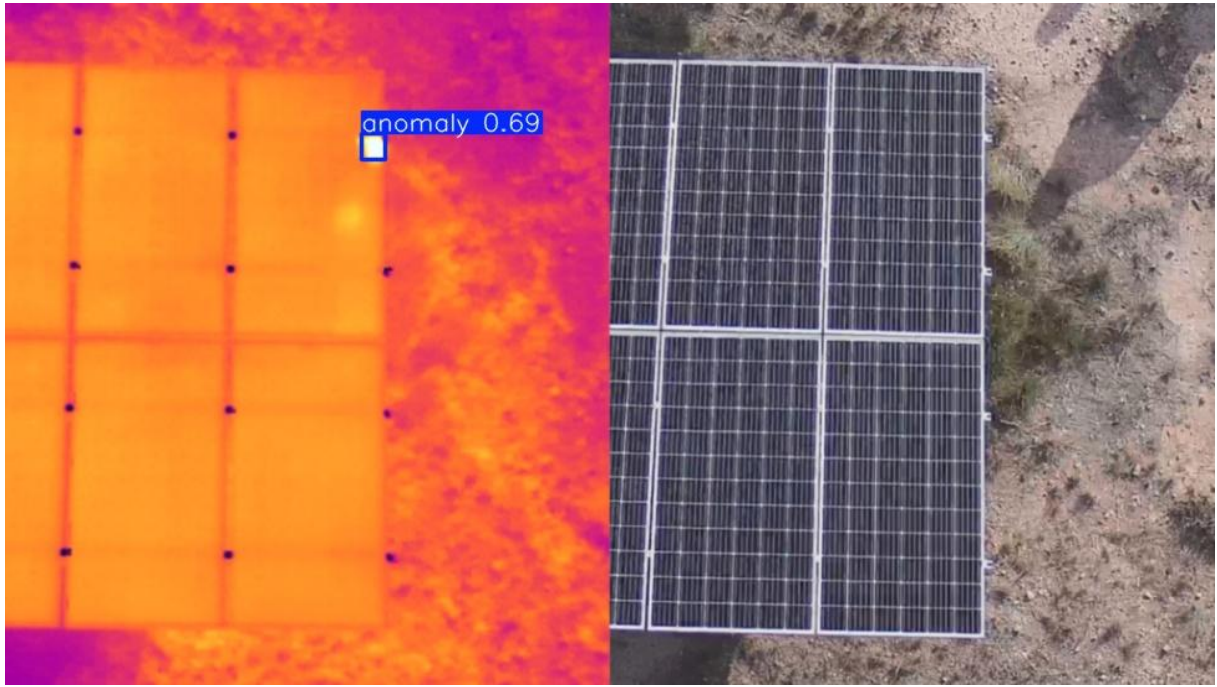

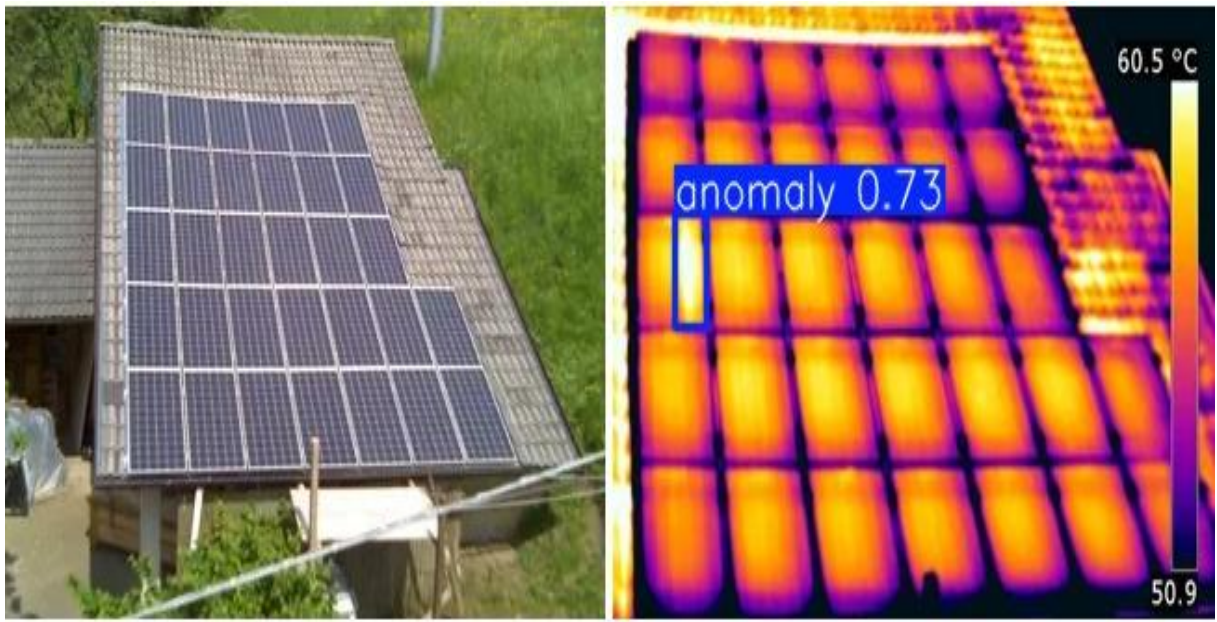

## Appendix F

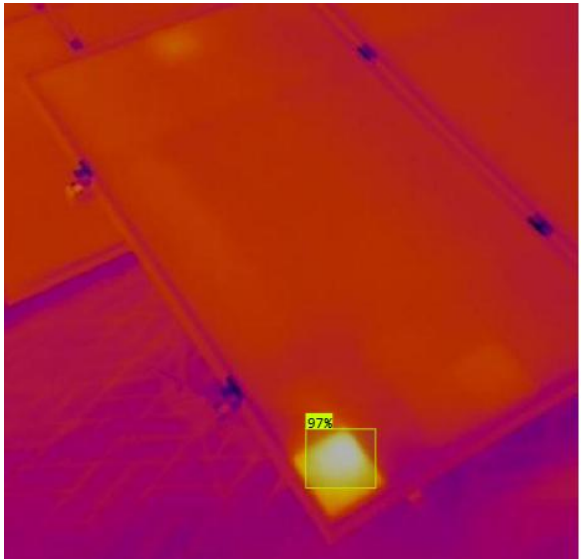
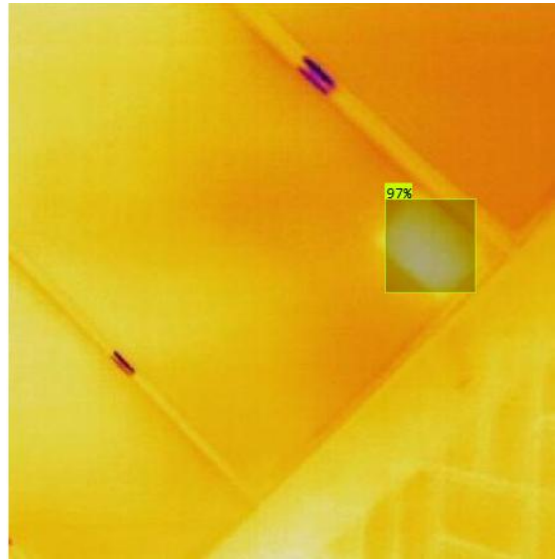
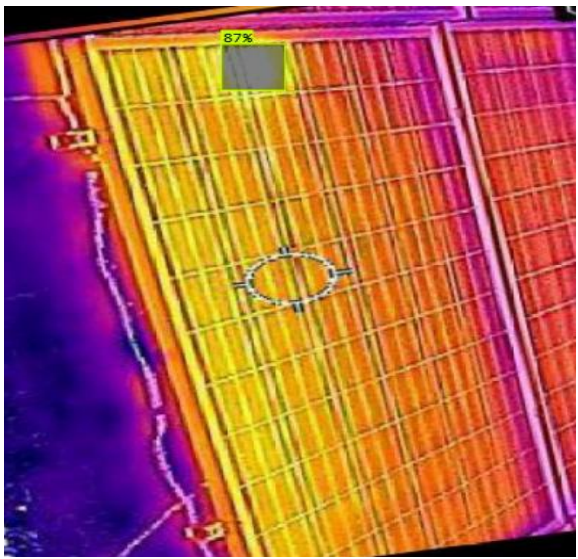
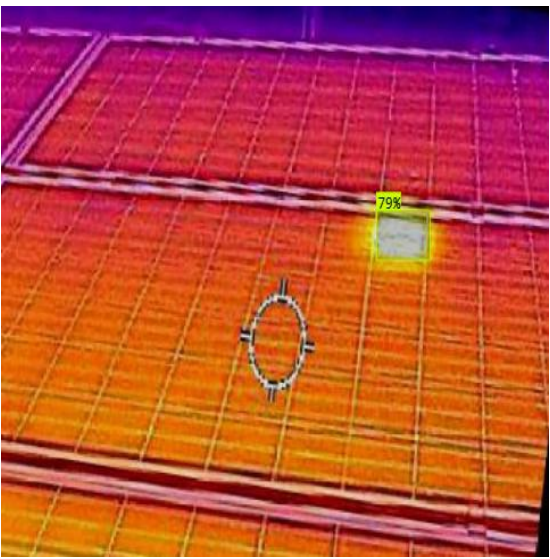
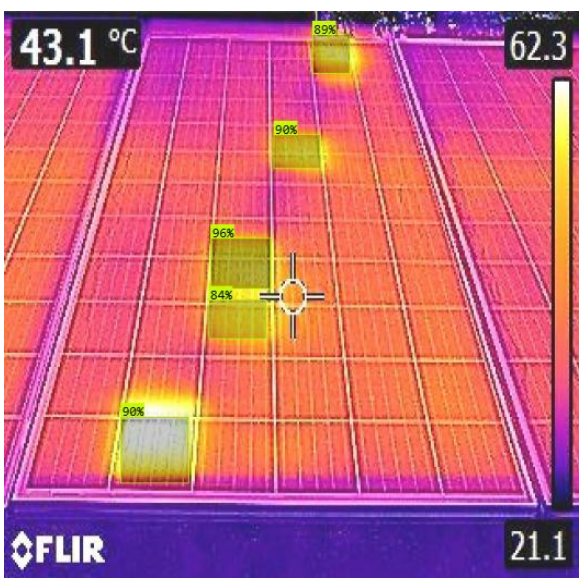
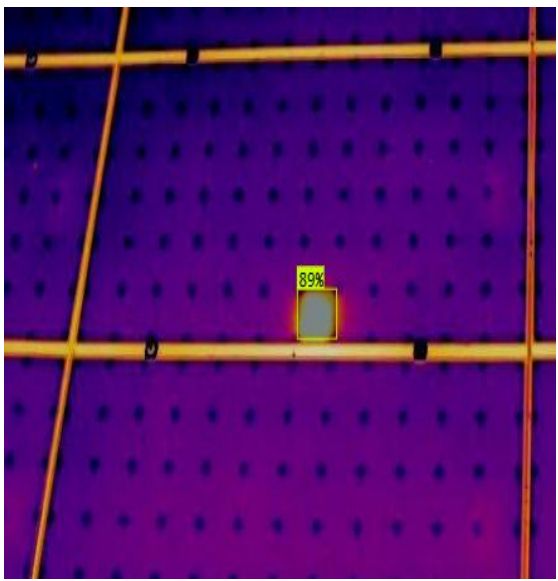

# References

[1]   Cubukcu, M. E. T. E., & Akanalci, A. (2020). Real-time inspection and determination methods of faults on photovoltaic power systems by thermal imaging in Turkey. Renewable Energy, 147, 1231-1238.

[2]   Zefri, Y., Sebari, I., Hajji, H., & Aniba, G. (2021). In-depth investigation of applied digital photogrammetry to imagery-based RGB and thermal infrared aerial inspection of large-scale photovoltaic installations. Remote Sensing Applications: Society and Environment, 23, 100576.

[3]   de Oliveira, A. K. V., Aghaei, M., & Rüther, R. (2022). Automatic inspection of photovoltaic power plants using aerial infrared thermography: a review. Energies, 15(6), 2055.

[4]   Pruthviraj, U., Kashyap, Y., Baxevanaki, E., & Kosmopoulos, P. (2023). Solar photovoltaic hotspot inspection using unmanned aerial vehicle thermal images at a solar field in south india. Remote Sensing, 15(7), 1914.

[5]   Buerhop, C., Bommes, L., Schlipf, J., Pickel, T., Fladung, A., & Peters, I. M. (2022). Infrared imaging of photovoltaic modules: a review of the state of the art and future challenges facing gigawatt photovoltaic power stations. Progress in Energy, 4(4), 042010.

[6]   Dhimish, M., & Badran, G. (2023). Investigating defects and annual degradation in UK solar PV installations through thermographic and electroluminescent surveys. npj Materials Degradation, 7(1), 14.

[7]   Rahaman, S. A., Urmee, T., & Parlevliet, D. A. (2022). Investigate the impact of environmental and operating conditions of infrared (IR) imaging on infrared thermography of PV modules to identify defects. Solar Energy, 245, 231-253.

[8]   Høiaas, I., Grujic, K., Imenes, A. G., Burud, I., Olsen, E., & Belbachir, N. (2022). Inspection and condition monitoring of large-scale photovoltaic power plants: A review of imaging technologies. *Renewable and Sustainable Energy Reviews*, *161*, 112353.

[9]   Yahya, Z., Imane, S., Hicham, H., Ghassane, A., & Safia, E. B. I. (2022). Applied imagery pattern recognition for photovoltaic modules' inspection: A review on methods, challenges and future development. Sustainable Energy Technologies and Assessments, 52, 102071.

[10]  Bosman, L. B., Leon-Salas, W. D., Hutzel, W., & Soto, E. A. (2020). PV system predictive maintenance: Challenges, current approaches, and opportunities. Energies, 13(6), 1398.

[11]  Hassan, S. and Dhimish, M., 2023. Dual spin max pooling convolutional neural network for solar cell crack detection. Scientific reports, 13(1), p.11099.

[12]  Drir, N., Mellit, A., Bettayeb, M. and Dhimish, M., 2025. Enhanced Photovoltaic Defect Detection Using Perceptual Loss in DCGAN and VGG16-Integrated Models on Electroluminescence Images. IEEE Journal of Photovoltaics.sssss

[13]  Herraiz, Á. H., Marugán, A. P., & Márquez, F. P. G. (2020). Photovoltaic plant condition monitoring using thermal images analysis by convolutional neural network-based structure. Renewable Energy, 153, 334-348.

[14]  Sapkota, R., Qureshi, R., Calero, M. F., Badjugar, C., Nepal, U., Poulose, A., ... & Karkee, M. (2024). YOLOv10 to Its Genesis: A Decadal and Comprehensive Review of The You Only Look Once (YOLO) Series. arXiv preprint arXiv:2406.19407.

[15]  Aktouf, L., Shivanna, Y., & Dhimish, M. (2024, November). High-Precision Defect Detection in Solar Cells Using YOLOv10 Deep Learning Model. In Solar (Vol. 4, No. 4, pp. 639-659). MDPI.

[16]  Hassan, S., & Dhimish, M. (2023, December). A Survey of CNN-Based Approaches for Crack Detection in Solar PV Modules: Current Trends and Future Directions. In Solar (Vol. 3, No. 4, pp. 663-683). MDPI.

[17]  Hussain, M., & Khanam, R. (2024, June). In-depth review of yolov1 to yolov10 variants for enhanced photovoltaic defect detection. In Solar (Vol. 4, No. 3, pp. 351-386). MDPI.

[18] Khanam, R., & Hussain, M. (2024). YOLOv11: An Overview of the Key Architectural Enhancements. arXiv preprint arXiv:2410.17725.

[19] Yang, X., Li, Y., Yang, L., Zhang, Y., Wang, X., & Zhang, Q. (2024). High-noise solar panel defect identification method based on the improved EfficientNet-V2. Journal of Renewable and Sustainable Energy, 16(5).

[20] Liu, B., Chen, L., Sun, K., Wang, X., & Zhao, J. (2024). A Hot Spot Identification Approach for Photovoltaic Module Based on Enhanced U-Net With Squeeze-and-Excitation and VGG19. IEEE Transactions on Instrumentation and Measurement.

[21] Jin, X., Xie, Y., Wei, X. S., Zhao, B. R., Chen, Z. M., & Tan, X. (2022). Delving deep into spatial pooling for squeeze-and-excitation networks. Pattern Recognition, 121, 108159.

[22] Qi, Q., Zhao, J., Lin, L., Zhang, X., & Tian, Y. (2024). Combined multi-level context aggregation and attention mechanism method for photovoltaic panel extraction from high resolution remote sensing images. International Journal of Remote Sensing, 45(11), 3560-3576.

[23] Gong, B., An, A., Shi, Y., & Zhang, X. (2024). Fast fault detection method for photovoltaic arrays with adaptive deep multiscale feature enhancement. Applied Energy, 353, 122071.

[24] Wang, Y., Hou, T., Zhang, X., Shangguan, H., Zhang, P., Li, J., & Wei, B. (2023). Surface defect detection of solar cell based on similarity non-maximum suppression mechanism. Signal, Image and Video Processing, 17(5), 2583-2593.

[25] Mazen, F. M. A., Seoud, R. A. A., & Shaker, Y. O. (2023). Deep learning for automatic defect detection in PV modules using electroluminescence images. IEEE Access, 11, 57783-57795.

[26] Mahasin, M., & Dewi, I. A. (2022). Comparison of CSPDarkNet53, CSPResNeXt-50, and EfficientNet-B0 backbones on YOLO v4 as object detector. International journal of engineering, science and information technology, 2(3), 64-72.

[27] Jiang, P., Ergu, D., Liu, F., Cai, Y., & Ma, B. (2022). A Review of Yolo algorithm developments. Procedia computer science, 199, 1066-1073.

[28] Wang, Z., Zhou, D., Guo, C., & Zhou, R. (2024). Yolo-global: a real-time target detector for mineral particles. Journal of Real-Time Image Processing, 21(3), 1-13.

[29] Dubey, S. R., Singh, S. K., & Chaudhuri, B. B. (2022). Activation functions in deep learning: A comprehensive survey and benchmark. Neurocomputing, 503, 92-108.

[30] Apicella, A., Donnarumma, F., Isgrò, F., & Prevete, R. (2021). A survey on modern trainable activation functions. Neural Networks, 138, 14-32.

[31] Ni, X., Ma, Z., Liu, J., Shi, B., & Liu, H. (2021). Attention network for rail surface defect detection via consistency of intersection-over-union (IoU)-guided center-point estimation. IEEE Transactions on Industrial Informatics, 18(3), 1694-1705.

[32] Soydaner, D. (2020). A comparison of optimization algorithms for deep learning. International Journal of Pattern Recognition and Artificial Intelligence, 34(13), 2052013.

[33] Schmidt, R. M., Schneider, F., & Hennig, P. (2021, July). Descending through a crowded valley-benchmarking deep learning optimizers. In International Conference on Machine Learning (pp. 9367-9376). PMLR.

[34] Jia, Y., Chen, G., & Zhao, L. (2024). Defect detection of photovoltaic modules based on improved VarifocalNet. Scientific Reports, 14(1), 15170.

[35] Xia, H., Yang, B., Li, Y., & Wang, B. (2022). An improved CenterNet model for insulator defect detection using aerial imagery. Sensors, 22(8), 2850.

[36] Wang, J., Zhang, R., & Zheng, X. (2023, May). Photovoltaic Panel Intelligent Detection Method Based on Improved Faster-RCNN. In 2023 IEEE 3rd International Conference on Electronic Technology, Communication and Information (ICETCI) (pp. 1565-1569). IEEE.

[37] Ultralytics YOLOv5 (2023). Link: https://docs.ultralytics.com/models/yolov5/#overview

[38]   Ultralytics YOLOv9 (2024). Link: https://docs.ultralytics.com/models/yolov9/#what-tasks-and-modes-does-yolov9-support

[39]   Ultralytics YOLOv11 (2024). Link: https://docs.ultralytics.com/models/yolo11/

[40]   Han, X., Wang, X., Chen, C., Li, G., & Piao, C. (2023). Hot Spot Detection of Thermal Infrared Image of Photovoltaic Power Station Based on Multi-Task Fusion. Journal of Information Processing Systems, 19(6), 791-802.

[41]   Numan, A., Hussein, H. A., & Dawood, Z. S. (2021). Hot spot analysis of photovoltaic module under partial shading conditions by using IR-imaging technology. Engineering and Technology Journal, 39(9), 1338-1344.

[42]   Ali, M. U., Saleem, S., Masood, H., Kallu, K. D., Masud, M., Alvi, M. J., & Zafar, A. (2022). Early hotspot detection in photovoltaic modules using color image descriptors: An infrared thermography study. International Journal of Energy Research, 46(2), 774-785.